\documentclass[a4paper,10pt,titlepage,twoside]{article}

\usepackage{ucs}
\usepackage[utf8x]{inputenc}
\usepackage{graphicx}
\usepackage{bm}

\usepackage[numbers]{natbib}
\usepackage{fancyhdr}
\author{\\ \bf{C\'{e}line Lichtensteiger, Pavlo Zubko and Jean-Marc Triscone}\\ \it{Universit\'{e} de Gen\`{e}ve}\\ \it{24 Quai Ernest Ansermet, CH-1211 Gen\`{e}ve 4, Switzerland}\\ \\ \bf{Massimiliano Stengel}\\ \it{Institut de Ci\`{e}nca de Materials de Barcelona (ICMAB-CSIC)}\\ \it{Campus UAB, E-08193 Bellaterra, Spain}\\ \\ \bf{Philippe Ghosez}\\ \it{Physique Théorique des Matériaux, Université de Li\`{e}ge}\\ \it{All\'{e}e du 6 Ao\^{u}t 17 (B5), B-4000 Sart Tilman, Belgium}\\ \\ \bf{Pablo Aguado-Puente and Javier Junquera}\\\it{Departamento de Ciencias de la Tierra y Fisica de la Materia Condensada}\\ \it{Universidad de Cantabria, Cantabria Campus Internacional}\\ \it{Avda. de los Castros s/n, E-39005 Santander, Spain}\\ \\ Ref: Chapter 12 in Oxide Ultrathin Films: Science and Technology (2012)\\ Edited by Gianfranco Pacchioni and Sergei Valeri\\ WILEY-VCH ISBN 978-3-527-33016-4 }
\title{\bf{Ferroelectricity in ultrathin film capacitors} }
\date{}

\newcommand{\bto}{{BaTiO$_3$}}

\newcommand{\etal}{{\it{et al.}}}

\begin{document}

\maketitle
\tableofcontents

\fancyhead[LE,RO]{\slshape\thepage}
\fancyhead[RE]{\slshape \leftmark}
\fancyhead[LO]{\slshape \rightmark}
\fancyfoot[LE]{\slshape Ferroelectricity in ultrathin film capacitors}
\fancyfoot[RE]{}
\fancyfoot[RO]{\slshape C. Lichtensteiger, P. Zubko, M. Stengel, P. Aguado-Puente, J.-M. Triscone, Ph. Ghosez, and J. Junquera}
\fancyfoot[LO]{}
\fancyfoot[C]{}

\section{Introduction}
\label{s_Introduction}

 Ninety years ago Valasek discovered that the permanent polarization
 of Rochelle salt could be reversed with an applied electric field, 
 giving rise to a hysteretic polarization-field 
 response~\cite{Valasek:1920p1926,Valasek:1921p2049}.
 By analogy with ferromagnets, such materials became known as ferroelectrics, and have since attracted considerable interest from a
 fundamental aspect and because of their wide range of
 potential applications. 
 Ferroelectric materials are also \emph{piezoelectric}, i.e. electric 
 charges appear at their surface (due to changes in polarization) 
 when they are under mechanical strain, and vice versa,
 as well as \emph{pyroelectric}, i.e. their electrical dipole moment 
 depends on the temperature. 
 These properties find many applications and make
 these materials technologically important. 
 Thermal infrared pyroelectric detectors,
 ultrasound transducers, non-volatile ferroelectric Random Access Memories
 (FeRAM)~\cite{Scott:2000p9969} for smart cards and 
 portable electronic devices, surface
 acoustic wave devices for filters in telecommunications and
 gravimetric sensors, membrane type actuators useful in micropump
 devices, ultrasonic micromotors, ultrasonic transducers and
 sensors for medical imaging, and other medical applications for
 blood pressure control are just a few examples of the many
 technological applications using these materials. 
 Going down to the limit of ultrathin films holds promise for a new generation of devices such as ferroelectric tunnel 
 junctions~\cite{Tsymbal-06, Maksymovych-09, Garcia-09, Garcia-10}
 or resistive memories~\cite{Kohlstedt-05}. 
 However, these length scales also make the devices sensitive to 
 parasitic effects related to miniaturization, 
 and a better understanding of what happens as size is reduced is of 
 practical importance for the future development of these devices. 
 This chapter is about what happens to ferroelectric films as we go nano. 
 
This is a particularly exciting time for nanoscale physics, as the experimental advances in materials preparation and characterization have come together with great progress in theoretical modeling of ferroelectrics, and both theorists and experimentalists can finally probe the same length and time scales. This allows realtime feedback between theory and experiment, with new discoveries now routinely made both in the laboratory and on the computer. Throughout this chapter, we will highlight the recent advances in density functional theory based modeling and the role it played in our understanding of ultrathin ferroelectrics. We will begin with a brief introduction to ferroelectricity and ferroelectric oxides in section~\ref{s_Ferroelectricity_and_oxides}, followed by an overview of the major theoretical developments in section~\ref{s_theoretical}. We will then discuss some of the subtleties of ferroelectricity in perovskite oxides in section~\ref{s_ferroinbulk}, before turning our attention to the main subject of the chapter  -- ferroelectricity in ultrathin films in section~\ref{s_external-parameters}. In this section we will discuss in detail the influence of the mechanical, electrical and chemical boundary conditions on the stability of the polar state in a parallel plate capacitor geometry, introducing the notion of depolarization fields that tend to destabilize ferroelectricity. In section~\ref{s_domains}, we will look at other ways in which a thin ferroelectric can preserve its polar state, focusing on ferroelectric domains and domain walls. Finally, in section~\ref{s_artificiallayer} we will briefly discuss artificially layered ferroelectrics and the potential they hold as tailor-made materials for electronic applications.

Of course, it is impossible to summarize 90 years of research within the scope of this chapter. We will only review some of the stimulating developments that took place in the field of ferroelectric thin films within the last few years and will point the attention of the interested reader to the
 most comprehensive books and reviews summarizing the state of the
 art of this exciting field of research, both theoretically and experimentally.

\section{Ferroelectricity. Basic definitions}
\label{s_Ferroelectricity_and_oxides}

\begin{figure}[!htb] 
    \begin{center}
       \includegraphics[width=12cm,keepaspectratio]{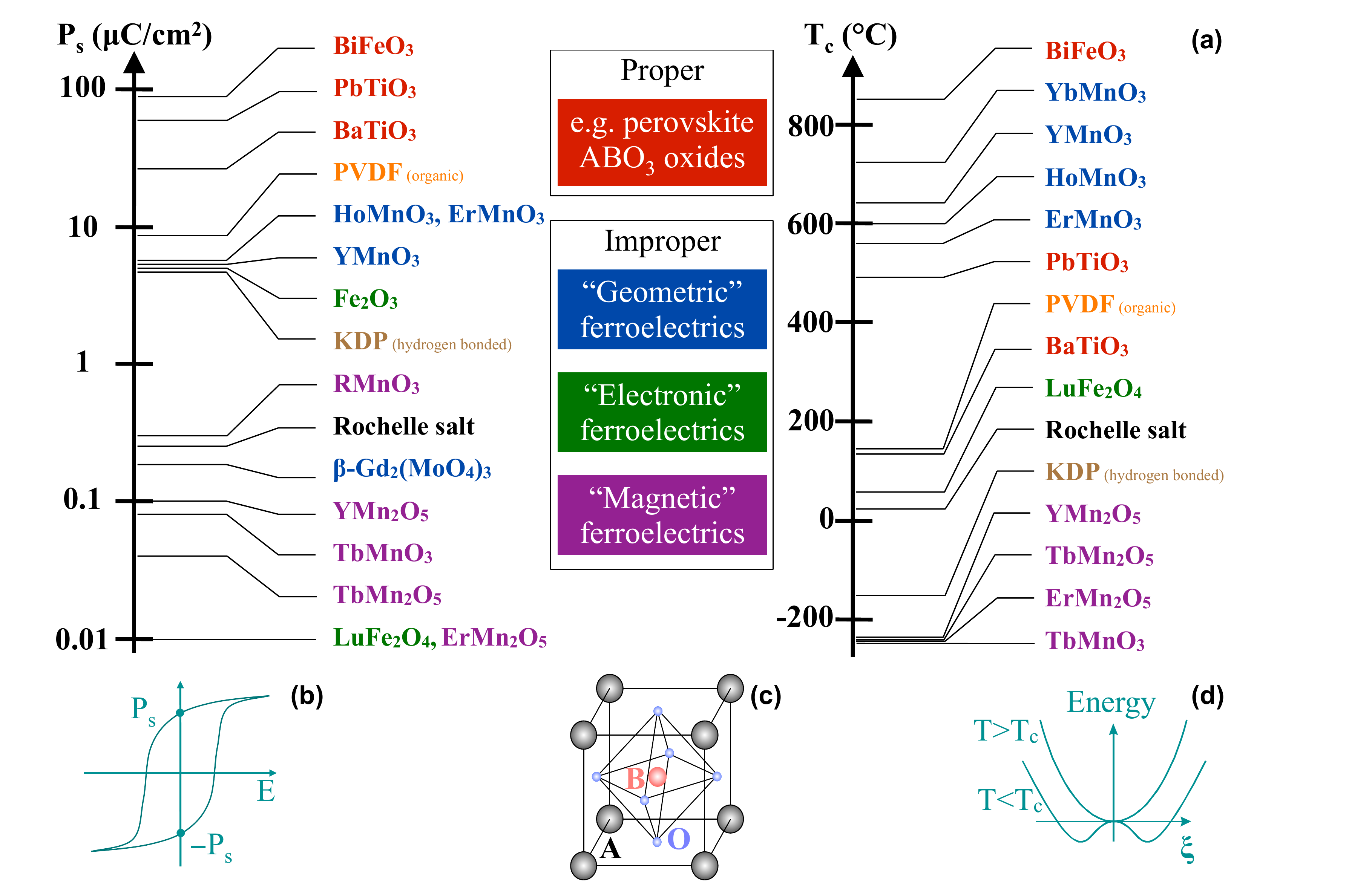}
       \caption{ Ferroelectrics encompass an enormous range of compounds, 
                 with a multitude of structures and compositions, 
                 both organic and inorganic. 
                 (a) Their defining and 
                 technologically relevant properties – the spontaneous 
                 polarization ($\bm{P}_{\rm s}$) and the critical
                 temperature ($T_{\rm c}$) – cover a wide
                 range of values. 
                 (b) Ferroelectric materials display an hysteresis loop between 
                 polarization and electric field.
                 (c) Among proper 
                 ferroelectrics, perovskite oxides with the generic formula 
                 ABO$_{3}$ have received most attention due to their simple structure, 
                 chemical stability and large polarization values. 
                 The structure of the high-temperature
                 paraelectric phase for most ferroelectric perovskite oxides
                 is cubic, as the one shown schematically.
                 (d) While hysteresis loops constitutes the measurement
                 of choice to experimentally demonstrate ferroelectricity,
                 the existence of a double-well energy landscape as a 
                 function of a macroscopic polar mode coordinate $\xi$ is 
                 usually considered as the theoretical fingerprint of the
                 ferroelectric instability. }
       \label{f:Polarization} 
    \end{center} 
 \end{figure}

\emph{Ferroelectrics} are materials exhibiting a spontaneous 
 electric polarization that can be switched by applying an electric field.
 The value of the spontaneous polarization can vary over several orders of magnitude depending on the material (see Fig~\ref{f:Polarization} (a)). The word ``ferroelectric'' is actually somewhat of a misnomer as these materials rarely contain iron.  The prefix ``ferro'' was instead adopted by analogy with the more mature field of ferromagnetism, which has many parallels with ferroelectricity:
  (i) a ferromagnet has a \emph{spontaneous magnetization} that can be switched
 by an external magnetic field,
 (ii) the switching process can be associated in both cases 
 with an hysteresis loop [see Fig.~\ref{f:Polarization} (b)],
 (iii) very often there is a coupling between polarization
 (either magnetic or electric) and the shape of the unit cell (strain),
 (iv) both the ferroelectric and ferromagnetic polarization decrease with
 increasing temperature up to a critical $T_{\rm c}$, 
 where a phase transition to 
 a high-symmetry unpolarized phase takes place and the corresponding
 polarization order parameter vanishes, and
 (v) even below $T_{\rm c}$ the macroscopic polarization
 might vanish if the homogeneously polarized state breaks into 
 \emph{domains} (regions with oppositely oriented polarization within 
 the sample).
 Despite all these similarities, one must keep in mind that the
 microscopic origin of these two phenomena are different,
 so a direct extrapolation from one world into the other is not
 always possible.
 As an example, the domain walls in ferromagnets are orders of magnitude
 wider than those in ferroelectrics.
 A complete side-by-side analysis of 
 analogies and differences can be found in Ref.~\cite{Spaldin-07}.
 
 When an electric field is applied to any \emph{insulating} material, 
 the bound electric charges inside the material will move on a short scale. 
 The material becomes polarized. 
 If the applied electric field $\bm{E}$ is not too large,
 then the polarization response 
 of the dielectric $\bm{P}_{\rm d}$ can be assumed to be linear,

 \begin{equation}
    \bm{P}_{\rm d}({\bm{E}}) = 
    \varepsilon_{0} \chi \bm{E}, 
    \label{eq:lindielpol}
 \end{equation}

 \noindent where $\varepsilon_{0}$ is the vacuum permittivity,
 and $\chi$ is the 
 electric susceptibility tensor of the dielectric
 (throughout this chapter we will use the SI system of units).
 If the medium is isotropic the $\chi$ can be considered
 as a constant of proportionality independent of the direction.

 In a ferroelectric 
 material the symmetry is spontaneously broken so,
 even in the absence of an external electric field,
 there are two or more polar states with a non-vanishing 
 spontaneous polarization, $\bm{P}_{\rm s}$.
 If a \emph{small} electric field is applied when the system is in
 one of these stable polarization states, the total polarization 
 of the ferroelectric reads

 \begin{equation}
    \bm{P}_{\rm f} (\bm{E}) = 
    \bm{P}_{\rm s} + \varepsilon_{0} \chi \bm{E}.
    \label{eq:ferropol}
 \end{equation}

 \noindent However, having an spontaneous polarization is just one
 of the conditions for a material to be classified as ferroelectric.
 The second requirement is that it should be possible to 
 switch from one polar state to another by 
 applying a field that is larger than a threshold value,
 called the \emph{coercive field}, $\bm{E}_{c}$.
 The switching mechanism typically proceeds via nucleation and growth of
 inverted domains~\cite{Shin-07}, with an external electric field
 inducing the domain-wall motion; this leads to the modification of
 the domain volumes and thus of the total polarization.
 Nevertheless, this is an area of continuing interest and controversy
 and there exist exceptions to the above scenario.
 For example, a continuous switching mechanism without domain formation has been reported very recently for PbTiO$_{3}$ thin films; note 
 that this behavior occurs only at certain values
 of thickness and temperature ~\cite{Highland-10}.

 Ferroelectricity has been reported in different families of compounds,
 including hydrogen bonded systems (for a review see Ref.~\cite{Koval-11} and the special issue
 in Volume 71 of Ferroelectrics journal, 1987) 
 such as KDP (potassium dihydrogen phospate; KH$_{2}$PO$_{4}$),
 polymeric systems (review in Ref.~\cite{Nalwa-95}) 
 such as PVDF (poly-vinylidene fluoride; 
 [-CH$_{2}$-CF$_{2}$-]$_{n}$),
 or the wide family of ferroelectric \emph{perovskites} ABO$_{3}$,
 where A is a mono-, di- or tri-valent cation
 and B is a penta-, tetra- or tri-valent cation, respectively.
 Perovskite ferroelectrics in particular have generated an enormous amount of interest, in part due to their simple structure, and will be the focus of this chapter. In most cases, this family of compounds crystallizes at high temperature in a simple cubic paraelectric phase with five atoms per unit cell~\footnote{ LiNbO$_{3}$ and related materials are exceptions, where
 the paraelectric phase has a rhombohedral unit cell with 
 two formula units per unit cell.}: the A cation at the corner of the cube, the B cation body-centered
 and the O anions face-centered, as represented in Fig.~\ref{f:Polarization}(c). This is known as the ``ideal'' perovskite structure and is
 typically taken as the high-symmetry reference configuration in 
 theoretical studies. Its simple crystalline structure along with a large spontaneous polarization makes this family of compounds
 widely used in applications and as model systems for theoretical studies.

 Below $T_{\rm c}$, the system undergoes a structural phase transition, 
 lowering the symmetry of the high-temperature paraelectric state.
 The unit cell becomes non-centrosymmetric, and can display 
 several equivalent configurations. 
 In the case of a tetragonal configuration, for example, there are two 
 equivalent antiparallel polarization states along the polar $c$-axis.
 In a simplified model, the phase transition can be
 characterized by the motion of the B-cations with respect to
 the O cage.
 This polar atomic distortion usually also induces a small
 deformation of the unit cell.

 A more detailed introduction of the essential background on 
 the physics of ferroelectrics can be found in the
 classical works of Lines and Glass~\cite{Lines-77},
 and Strukov and Levanyuk~\cite{Strukov:1998p9262}, 
 while a more recent perspective is available in 
 the first chapter of the book edited by Rabe, Ahn,
 and Triscone~\cite{Rabe-07.2}.

\section{Theoretical methods for the study of bulk ferroelectric materials}
\label{s_theoretical}

\subsection{Devonshire-Ginzburg-Landau (DGL) phenomenological theory}
\label{s_DGL}

 For many years, the method of choice for the theoretical study
 of ferroelectric materials has been the phenomenological approach
 based on the Ginzburg-Landau theory, first applied to the case
 of ferroelectrics by Devonshire~\cite{Devonshire-49,Devonshire-51}.
 The starting point of this theoretical model 
 of phase transitions is the identification of an {\em order parameter}:
 a physical quantity that is zero in the high-symmetry phase and 
 changes to a finite value once the symmetry is lowered.
 In the case of ferroelectric materials at the paraelectric-ferroelectric
 transition, this might be
 the polarization $\mathbf P$ or the electric displacement $\mathbf D$
 (as done in Chapter 3 of the classical work by 
 Lines and Glass, Ref.~\cite{Lines-77}).
 The second issue to deal with is the identification of the other relevant
 degrees of freedom in the problem under consideration.
 Independent variables have to be chosen among the conjugate pairs
 temperature-entropy ($T$,$S$) and stress-strain ($\sigma$-$\eta$).
 The choice is guided by the parameters under control in a given experiment.
 Depending on the combination of the degrees of freedom, the 
 corresponding thermodynamic potential is known under different names
 [Helmotz free energy, Gibbs free energy (elastic or electric), etc.,
 see Chapter 3 of Ref.~\cite{Lines-77}].
 It is important to note here that there is no ``best'' thermodynamic
 potential but only the ``most suitable'' functional for a specified 
 choice of the boundary conditions.

 The central {\it Ansatz} of the Landau approach is that the free
 energy can be represented as a Taylor expansion of the order parameter
 and the dependent variables
 in the neighborhood of the transition point,
 where only symmetry-compatible terms are retained.
 To illustrate the form of this functional in the simplest possible
 scenario, we can write the free energy of the ferroelectric in terms
 of a single component of the polarization, ignoring the strain field 
 and taking as zero the energy of the free unpolarized crystal, 

 \begin{equation}
   \mathcal{F} (P) = \frac{1}{2} \alpha_2 P^2 +
                     \frac{1}{4} \alpha_4 P^4 +
                     \frac{1}{6} \alpha_6 P^6 + \cdots
   \label{eq:landau-expansion}
 \end{equation}

 As previously mentioned, the coefficients of the expansion generally 
 depend on temperature. However, for practical purposes, most of
 them are usually considered as constants.  For instance, to deal
 with second order phase transitions, it is enough to  consider the
 quadratic coefficient as $T$-dependent, while $\alpha_{4}$ and $\alpha_{6}$
 are fixed to positive constant values. It is assumed that $\alpha_{2}$ has a
 linear dependency of the form $\alpha_{2} = \beta \left(T-T_{\rm{c}}\right)$
 with $\beta$ a positive constant and $T_{\rm{c}}$ the temperature of
 the ferroelectric-to-paraelectric phase transition.
 For $T>T_{\rm{c}}$ the quadratic term is positive and 
 the thermodynamic functional
 is a single well with its minimum corresponding to the non-polar
 $\bm{P} = 0$ phase.  For $T < T_{\rm{c}}$ the quadratic coefficient
 is negative, and the
 thermodynamic potential displays the typical double well shape.
 There are two minima for $\bm{P} \ne 0$ [see Fig.~\ref{f:Polarization}(d)].

 The coefficients of the expansion are fitted to the experiment,
 which is usually performed in a regime close to the phase transition
 (although lately some efforts to determine them from 
 first-principles~\cite{Iniguez-01.2} and thermodynamic 
 integrations~\cite{Geneste-09} have been carried out).
 Finally, the state of the system at a given temperature is found  
 by minimizing this free energy.
 We encourage the interested reader to pursue more on this subject in 
 the primer by Chandra and Littlewood~\cite{Chandra-07}.

 Despite the many virtues of DGL theory, 
 we must nevertheless keep in mind its limits of validity:
 it is strictly a macroscopic approach and thus it cannot describe 
 microscopic quantities such as atomic displacements.
 Therefore, it is expected to be valid only on length scales that are
 much larger than the lattice constant;
 some care has thus to be taken when extending DGL theory to the case of 
 ultrathin films.
 In addition, this model is only as good as its input parameters are,
 and extrapolations to temperature and strain values that lie outside
 a neighborhood of the phase transition should again be done with 
 caution.

\subsection{First-principles simulations}
\label{s_first-principles}

 As mentioned in the introduction, the last few years have witnessed 
 a rapid evolution of the atomistic modeling of materials.
 Nowadays, it is possible to describe and predict
 very accurately the properties of ferroelectrics using methods directly
 based on the fundamental laws of quantum mechanics and electrostatics.
 Even if the study of complex systems requires some practical approximations,
 these methods are free of empirically adjustable parameters.
 For this reason, they are referred to as  ``first-principles''
 or ``{\it ab initio}'' techniques.
 Due to its accuracy and efficiency, density functional theory (DFT) 
 has emerged as one of the most widespread methodological tools.
 A comprehensive overview of DFT,
 whose development earned W. Kohn his Nobel Prize in Chemistry in 1998,
 can be found in the excellent
 books by Richard M. Martin~\cite{Martin} or J. Kohanoff~\cite{Kohanoff}.

 First-principles simulations could not be applied to the study
 of ferroelectric materials until the early nineties, due to the lack
 of a formal theory of the polarization in periodic solids.
 While the polarization can easily be expressed in terms of the 
 charge distribution for molecules (finite systems),
 it cannot be obtained that way for crystals
 (infinite systems treated periodically). 
Indeed, the naive Clausius-Mossotti 
definition of the polarization as a dipole moment per unit volume cannot be used, as it depends on the (arbitrary) choice of the unit cell~\cite{Martin-74}. The solution to this long-standing problem appeared in the early 1990s, 
 and is often referred to as the ``modern theory of 
 polarization''~\cite{King-Smith-93}. 
 The basic idea
 is to consider the change in polarization~\cite{Resta-92} of a crystal as
 it undergoes some slow change, e.g. a displacement
 of one sublattice relative to the others, and relate it to
 the {\it current} that flows through the crystal during this adiabatic
 evolution of the system~\cite{Resta-94}.
 A review on the modern approach to the theory of polarization
 can be found in Refs.~\cite{Resta-07,Resta-10.1}.
 Once the access to the polarization of a periodic solid is granted, we can 
 compute relevant quantities of interest such as the Born effective charge defined in Eq.~(\ref{eq_born}), which are central to the phenomenon 
 of ferroelectricity as will be explained in Sec.~\ref{s_ferroinbulk}.

 After the establishment of the modern theory of polarization, a number
 of further methodological developments were proposed that 
 significantly enhanced the capabilities of first-principles simulations 
 of ferroelectric systems by allowing full control over
 the macroscopic electrical variables in periodic insulators.
 At the core of all these advances lies the method to introduce 
 a finite macroscopic
 electric field in a periodic quantum-mechanical 
 simulation~\cite{Souza-02,Umari-02}.
 Based on this method, rigorous strategies to perform 
 calculations at a specified 
 value of $\bm{P}$~\cite{Dieguez-06} or, later, of 
 the electric displacement 
 $\mathbf{D}$~\cite{Stengel-09-np} were subsequently established.
 These latter techniques opened the way to a direct computation 
 of the ``electrical
 equation of state'' (energy as a function of $\bm{P}$ or $\mathbf{D}$) 
 of a bulk ferroelectric material from first principles.
 This has several advantages, that we summarize hereafter.

 First, performing calculations at specified $\bm{P}$ or $\mathbf{D}$ 
 bears a direct relationship 
 to DGL theories, where $\bm{P}$ or $\mathbf{D}$ are commonly used as 
 independent electrical variable.
 Second, the strength of the polar instability can be directly 
 quantified by means
 of a generalized \emph{inverse permittivity}, 
 that is the second derivative of the
 total energy with respect to $\mathbf{D}$. This is an interesting alternative to the previously available estimates in terms of the imaginary frequencies of the soft mode.
 Third, by mapping the ground-state properties as a function 
 of $\mathbf{D}$ (or $\bm{P}$) one can
 very easily access all response properties of the system 
 (e.g. dielectric, piezoelectric, etc.) in a non-perturbative way 
 (i.e. non-linear at any order); this might be cumbersome 
 or impossible to obtain within the standard linear-response technique.
 There are further advantages related to the use of $\mathbf{D}$ 
 as fundamental electrical variables
 in the calculations of capacitors or superlattice geometries; 
 we shall review those
 later in the appropriate context.

 First-principles methods generate a wealth of microscopic information,
 reproducing the various ground state structures and functional properties
 of ferroelectric materials.
 The price to pay for this is its computational costs that thwarts
 their use in systems larger than a few hundreds of atoms or in 
 molecular dynamics simulations beyond a few tens of picoseconds.

\subsection{Second-principles methods: model Hamiltonians and shell models}
\label{s_second-principles}

 In order to extend the range of applicability of first-principles simulations
 to larger systems and time scales, and provide access to finite-temperature 
 thermodynamic properties,
 many efforts have been devoted to the development of methods that
 capture the essential physics with few parameters that can
 be directly extracted from DFT calculations.
 Among these ``second-principles'' methods, the most widespread
 ones are the model Hamiltonians and the shell models.
  
 The \emph{model (or effective) Hamiltonian} is a microscopic approach based on 
 (i) the identication of the most important degrees of freedom
 for describing the transition through the local mode approximation
 of Lines~\cite{Lines-69}, and (ii) to perform a low-order expansion
 of the energy in terms of these degrees of freedom, with coefficients
 directly determined from total energy DFT calculations. In the 1990's
 this model was successfully generalized to ferroelectric perovskite
 oxides~\cite{ZhongW-94.2,ZhongW-95.1}.
 In simple ferroelectrics, the first step greatly reduces the number
 of relevant degrees of freedom. For instance, in the prototype BaTiO$_{3}$
 ferroelectric oxide, the
 structural distortion nearly exactly
 corresponds to the freezing into the structure of one of the
 transverse optic modes at the $\Gamma$ point, usually referred
 to as the \emph{soft-mode}, and to a subsequent
 strain relaxation.  
 A reasonable approximation for the study of
 the phase transition is therefore to only consider explicitly
 (i) the ionic degree of freedom of the soft-mode by means of a
 local mode (i.e. a local cooperative pattern of atomic
 displacement) ${\bf{\xi}}_i$ that will be associated to each unit cell $i$ and
 (ii) the strains (homogeneous or inhomogeneous).
 As the amplitude of the ferroelectric
 distortion is relatively small, the Hamiltonian can then be
 written as a Taylor expansion in terms of ${\bf{\xi}}_i$ and
 strains limited to low orders (again, only the symmetry-allowed terms
 are kept).
 Once the expression for the model Hamiltonian is known, and the parameters
 have been fitted, one can very efficiently predict the value of the energy 
 for arbitrary configurations of the local cooperative patterns ${\bf{\xi}}_i$.
 The sampling of the parameter space can be performed via classical Monte-Carlo
 or Molecular Dynamic simulations to investigate the temperature behaviour
 of ferroelectrics.
 The standard output provides the mean
 values $< {\bf \xi}>$ and $<\eta>$ 
 (where $\eta$ stands for the homogeneous strain) 
 in terms of temperature, external
 pressure and electric field. The macroscopic polarization of the
 crystal $< \bm{P} >$  is also readily accessible since the local
 polarization ${\mathbf P}_i$  is directly proportional to the amplitude
 of the local mode ${\bf \xi}_i$,

 \begin{eqnarray}
 {\mathbf P}_i = \frac{1}{\Omega_0} \bar{Z}^* {\bf \xi}_i,
 \end{eqnarray}

 \noindent where $\bar{Z}^*$ is the Born effective charge of the local
 mode as defined later in Eq.~\ref{eq_born}, and $\Omega_0$ the unit cell volume.

 A step by step explanation about how to develop a model Hamiltonian can
 be found directly in the original paper by Zhong and 
 coworkers~\cite{ZhongW-95.1}.
 
 Within the \emph{shell model} a given ion of static
 charge $Z$ is modeled as a massive core, of mass $m$ and 
 charge $X$, linked to a massless shell of charge $Y=Z-X$.
 The core and the electronic shell within an ion are connected through a spring 
 of force constant $k$.
 This core-shell interaction might be considered anisotropic
 and anharmonic (containing up to fourth order terms)
 for the O$^{2-}$ ions to emphasize the large anisotropic polarization 
 effects at the oxygens associated with the ferroelectric distortions.
 In addition to the Coulombic interaction, the model contains
 pairwise short range potentials acounting for the effect of the 
 exchange repulsion between atoms.
 All these material-specific parameters are determined by 
 first-principles calculations.
 Then, the equations of motion are solved for all these coupled springs.
 A complete description of the method is available in Ref.~\cite{Sepliarsky-05}.
 This methodology is able to describe very well the phase behaviour 
 and ferroelectric properties of KNbO$_{3}$~\cite{Sepliarsky-00}, 
 PbTiO$_{3}$ (with surface effects~\cite{Sepliarsky-05.2}, 
 and the coupling with the substrate~\cite{Sepliarsky-06}
 or depolarization fields~\cite{Stachiotti-11}),
 BaTiO$_{3}$ and SrTiO$_{3}$~\cite{Tinte-04},
 including its solid solutions and ultrathin films~\cite{Tinte-01}.

\begin{figure}[!htb] 
    \begin{center}
       \includegraphics[width=10cm,keepaspectratio]{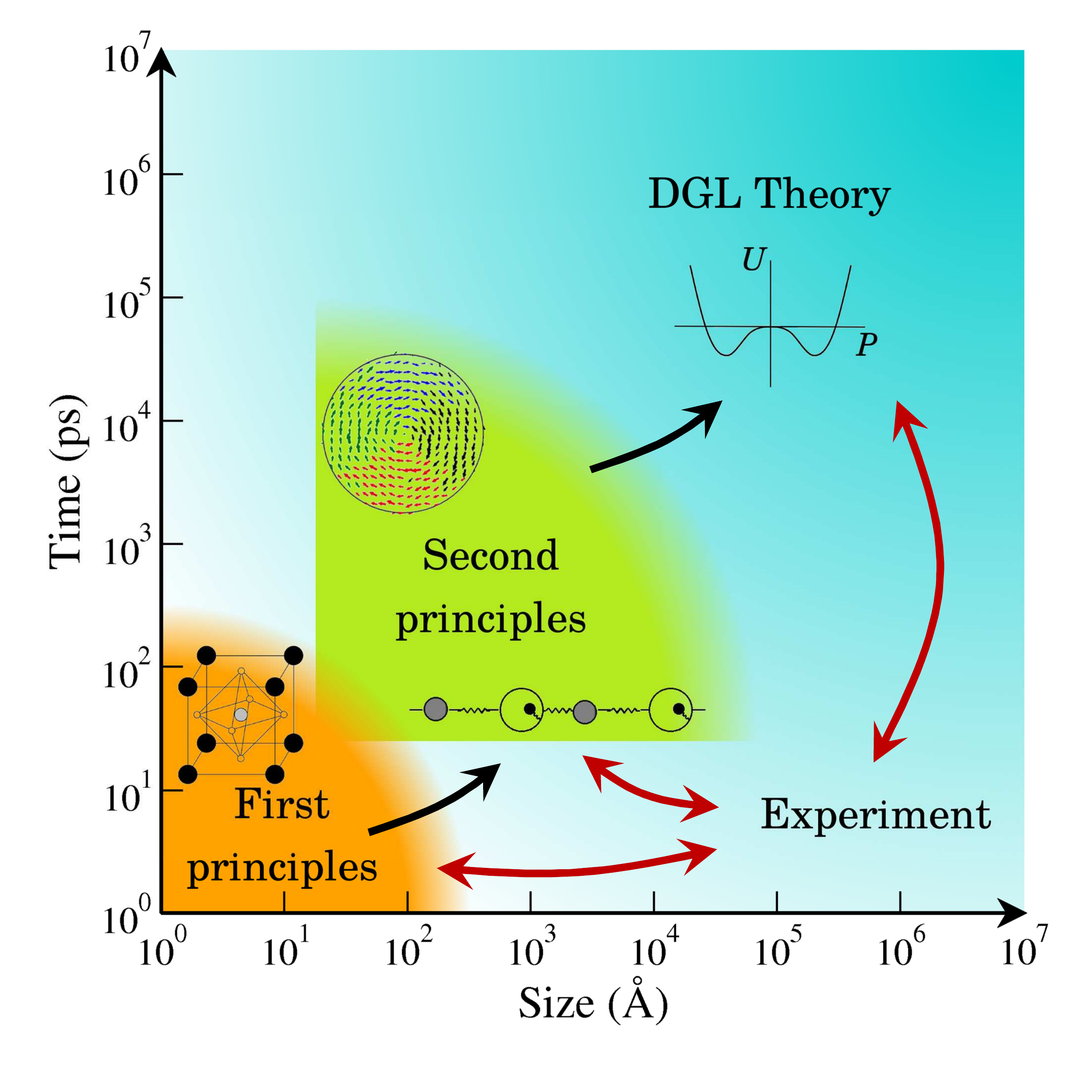}
       \caption{ Sketch with the different length and time scales accessible
                 with the various theoretical 
                 schemes presented in this chapter.
                 Black arrows indicate the interconnection of the methods.
                 First-principles methods with atomic resolution 
                 (represented by the balls and sticks cartoon),
                 feed second-principles models where only some degrees
                 of freedom are considered
                 (for instance, the soft mode in every unit cell,
                 represented by the arrows or the springs in the cartoons).
                 Parameters for the phenomenological 
                 Devonshire-Ginzburg-Landau methods
                 can be determined from atomistic methods. The arrows in red stress the interconnection between experiments and theories at the different levels.}
       \label{f:theory-methods}
       \end{center}
 \end{figure}

 As a summary of this Section, we schematically represent in Fig.~\ref{f:theory-methods} the length and time scales accessible within the different
 methodologies discussed before.

\section{Modeling ferroelectricity in oxides}
\label{s_ferroinbulk}

 Since the early 1990s, many density-functional theory calculations 
 were carried out, as a complement to the experiment, 
 in order to understand the microscopic mechanisms that are
 responsible for ferroelectric phase transitions, and to
 provide a firmer theoretical basis to existing classical theories. 

 Special attention was given to the interplay between the electronic 
 and lattice-dynamical properties, with the intention
 of clarifying the physical nature of the structural instability.
 The classical groundbreaking work along these lines is due to 
 Cochran~\cite{Cochran-60}, who first realized the importance
 of lattice dynamics in describing the ferroelectric distortion,
 and established the important concept of \emph{soft mode}
 as the basic ingredient in the theory.

 The explanation focuses on a special class of ABO$_{3}$ ferroelectric 
 oxides: those  where the B-site 
 cation is formally in a $d^0$ oxidation state (common examples are 
 BaTiO$_3$, KNbO$_3$). 
 This means that ``true'' $d$ electrons are, in principle, absent from the 
 compound. Nevertheless, the occupied O ($2p$) and the empty B ($d$) orbitals 
 partially hybridize, producing a bonding of mixed ionic-covalent 
 nature~\cite{Cohen-92,Posternak-94}.
 A key observation here is that these hybridizations are very sensitive to 
 the O-B distance. Based on this remark, Harrison~\cite{Harrison-80}
 predicted a dynamical transfer of electronic charge when the atoms are displaced.

 To understand qualitatively why this happens, it is useful to 
 introduce the concept of \emph{Born effective charge}, i.e. the polarization 
 linearly induced by a small displacement of the atom $\alpha$  within zero 
 macroscopic electric field~\cite{Ghosez-98}.
 For the sake of simplicity, if we consider a one-dimensional system
 it can be written as

 \begin{equation}
    Z^*_\alpha = \frac{\partial P}{\partial x_\alpha} \Big|_{\bm{E}=0}
    \label{eq_born}
 \end{equation}

\begin{figure}[!htb] 
    \begin{center}
       \includegraphics[width=12cm,keepaspectratio]{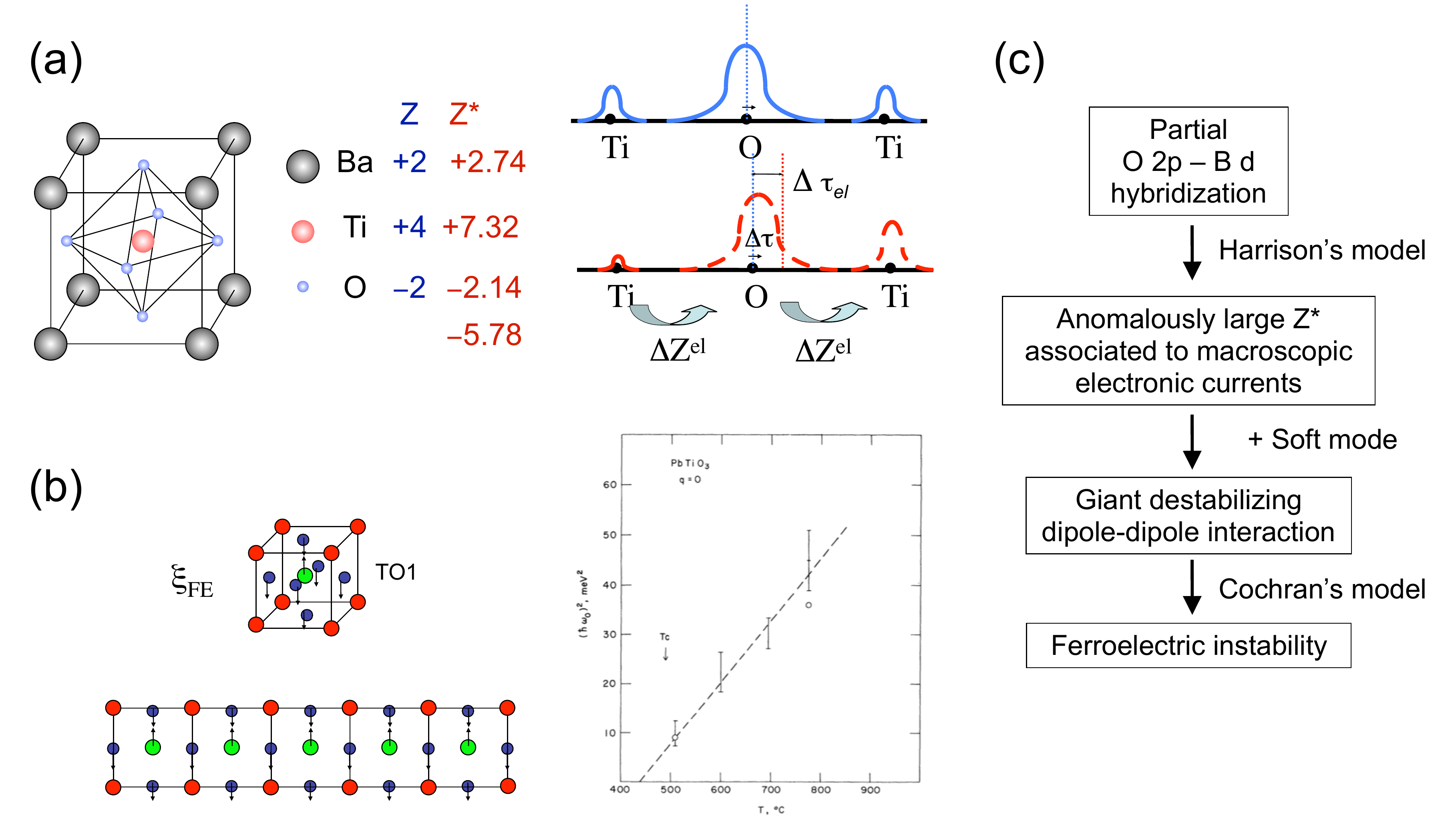}
       \caption{ (a) The concept of Born effective charge $Z^*$ illustrated for the case of BaTiO$_3$. $Z_\alpha^*$ is equal to the polarization linearly induced by a small displacement of each atom $\alpha$ within zero macroscopic electric field~\cite{Ghosez-98}. In a purely ionic system, the Born charges coincide with the static charges $Z$, while in a real material the Born charges account for non-trivial electronic effects as well~\cite{Resta-07}.  The Ti-O hybridization in BaTiO$_3$ implies that the electronic valence orbitals do not follow rigidly the atom they belong to, but also significantly \emph{polarize} in response to the displacement. As a consequence the Born charges are strongly ``anomalous''~\cite{ZhongW-94,Resta-99}, i.e. a displacement of a given atomic sublattice gives rise to a spontaneous polarization that is much larger than that predicted by the static nominal charges. (b) Illustration of the ionic displacements corresponding to the soft mode. The structural distortion ${\bf{\xi}}_{FE}$ is represented in a perovskite unit cell and corresponds to the displacement of the $B$ and $O$ atoms with respect to the A atoms. The local cooperative pattern of such atomic displacements corresponds to one of the transverse optic modes and is usually referred to as the \emph{soft-mode}, as the frequency $\omega$ associated to this mode will become imaginary (so that $\omega^2<0$ ), as illustrated by the experiment on PbTiO$_3$ from ref.~\cite{Shirane-70}. (c) Flowchart summarizing the connection between the O $2p$--B $d$ cation hybridizations and the ferroelectric instability of ABO$_{3}$ ferroelectric perovskite oxides.}

       \label{f:ferrobulk}
    \end{center}
 \end{figure}

 \noindent Within an extreme rigid-ion model, the Born charges coincide
 with the static charge of the model ion (the ``nominal'' value),
 while in a real material the Born charges account for electronic
 effects as well~\cite{Resta-07}.
 In the case of ABO$_{3}$ ferroelectrics, for example, the aforementioned O-B 
 hybridization implies that the electronic valence orbitals do not follow 
 rigidly the atom they belong to, but also significantly \emph{polarize} in 
 response to the displacement.
 As a consequence the Born charges are strongly
 ``anomalous''~\cite{ZhongW-94,Resta-99} as shown in Fig.~\ref{f:ferrobulk}(a) for BaTiO$_3$, i.e. a displacement of 
 a given atomic sublattice gives rise to a spontaneous polarization
 that is much larger than that predicted by the static nominal charges.
 A first-principles based band-by-band decomposition 
 clearly demonstrates the link between the
 O $2p$-B $d$ hybridization and the giant dynamical 
 charges~\cite{Ghosez-95,Ghosez-00.2}.
 
 The anomalously large Born effective charges have profound consequences
 on the lattice dynamics. The relevant phonons here are the transverse modes 
 at the zone center ($\Gamma$ point), where all equivalent atoms in every 
 unit cell are displaced by the same amount [see Fig.~\ref{f:ferrobulk}(b)].
 Following the original idea of Cochran, the leading contributions to 
 the phonon frequencies can be separated into short-range and long-range
 as follows.
 A giant dynamical charge implies a large
 long-range dipole-dipole interaction, which in turn tends to ``soften'' 
 the optical phonons of the crystal where the O and B cations move
 in antiphase. 
 In addition to this electrostatic term (which destabilizes the
 centrosymmetric configuration and thus favors ferroelectricity)
 we have the short-range repulsion between the ionic 
 cores, which has the opposite effect.
 The stability or instability of the ferroelectric mode is ultimately decided 
 by the delicate compensation of stabilizing short-range 
 forces and destabilizing dipole-dipole interactions (see Fig. 1 
 of Ref.~\cite{Ghosez-96}).  
 Whenever the latter prevail, the frequency $\omega$ of the lowest 
 transverse optical mode becomes imaginary, so that $\omega^2<0$, 
 which reflects the negative curvature at the origin
 of the total energy surface as a function of the mode amplitude 
 (this is the origin of the term \emph{soft mode}). 
 As this delicate balance between short- and long-range forces is 
 very sensitive to small modifications of the volume or of the Born 
 effective charges~\cite{Ghosez-96},
 the strength (or the very presence) of the instability can vary 
 greatly from one perovskite to another.
 This can be taken as a first-principles confirmation of the 
 old hypothesis formulated by Slater~\cite{Slater-50} 
 who suggested that the ferroelectric behaviour of BaTiO$_{3}$
 might be caused by long-range dipolar forces that (via
 the Lorentz local effective field) would destabilize the 
 high-symmetry configuration (the latter would be stabilized by 
 more localized forces).
 The softening of a transverse optic mode has been experimentally 
 confirmed by spectroscopic techniques~\cite{Harada-71,Scott-74}.

 In the particular case of BaTiO$_{3}$ discussed above, which is the 
 paradigmatic example of a ferroelectric perovskite oxide,
 both the short-range and dipolar parts of the interatomic interactions 
 are highly anysotropic, and therefore the correlation of the atomic 
 dispacements is much stronger along the Ti-O chains.
This concept of ``chain instability'' also applies to other materials like bulk KNbO$_{3}$.
 Other compounds like e.g. PbTiO$_{3}$ present significant differences
 and deserve a separate discussion. For example, the A-site cation in
 both BaTiO$_{3}$ and KNbO$_{3}$ is usually considered to be chemically 
 inert; conversely, Pb has a significant covalent interaction with 
 oxygen. As a consequence, the correlations
 of the atomic displacements in PbTiO$_3$ appear to be much more isotropic. 
 More importantly, the substantial role played by the A-site cation in the 
 ferroelectric instability of Pb-based (and Bi-based) compounds in practice
 lifts the traditionally assumed requirement of $d^0$-ness in the B-site 
 cation. 
 This observation opened new avenues of research in the context of multiferroics
 materials, where $d$ electrons are essential or the compound will not be 
 magnetic~\cite{Nicola-00}.
 In recent years, many different mechanisms for ferroelectricity emerged, that
 depart even more drastically from the original soft-mode picture by Cochran;
 this is a very lively and active area of research, whose coverage goes
 unfortunately beyond the scopes and size limitations of the present work.
 
 In summary, the driving force of conventional ferroelectric transitions 
 can be understood in terms of the anomalously large Born effective charges, 
 that couple to the atomic displacements via a giant dipolar interaction
 and produce an unstable ``soft'' phonon mode [see Fig.~\ref{f:ferrobulk}].
 The large anomalous charges, in turn, emerge from the hybridization 
 between O $2p$ and B-cation $d$ orbitals, which is responsible
 for the polarization of the valence orbitals upon atomic displacement.

 It is important to emphasize the \emph{cooperative} nature of the
 ferroelectric phenomena: a single atomic displacement is not enough
 to trigger the ferroelectric distortion in a crystal.
 Only the cooperative displacement of a group of atoms inside
 the \emph{correlation volume} is enough to induce the ferroelectric
 instability~\cite{Geneste-08}. It is natural then to expect that this delicate balance will be modified in thin films, where the correlation volume is truncated. We shall address this point in the following section.

\section{Theory of ferroelectric thin films}
\label{s_external-parameters}

 As discussed above, ferroelectricity can be 
 in many cases linked to spontaneous atomic off-center displacements, 
 resulting from a delicate balance between long-range dipole-dipole 
 Coulomb interaction and short-range covalent repulsions.
 In ultrathin films and nanostructures, both interactions are modified 
 with respect to the bulk.
 Short-range interactions are modified at surfaces and interfaces,
 due to the different chemical environment (chemical boundary conditions).
 Long range interactions are truncated 
 due to lack of periodicity and is strongly dependent
 on the electrical boundary conditions. 
 As stated above, the ferroelectric instability is also strongly
 sensitive to strain and will be influenced by mechanical boundary
 conditions such as epitaxial strains. These different
 factors can act independently to either enhance or suppress
 ferroelectricity.

 For many years, samples below a certain
 size did not display ferroelectricity.
 It turned out that the reason for this suppression was not 
 because of intrinsic size effects, but rather due to difficulties in
 fabrication. The fact that the experimentally obtained minimum thickness for a
 ferroelectric thin film has decreased by orders of magnitude over
 the years is a clear sign that for the most part the suppression was
 due to limitations in sample quality. 
 For example, dead layers (incomplete screening),
 misfit dislocations and partial relaxation of the lattice, 
 grain boundaries and defects such as oxygen vacancies are all known
 to strongly influence the ferroelectric properties.

 In what follows we briefly summarize the most important concepts behind those
 different boundary conditions and their influence on the ferroelectric 
 properties.
 For a more complete and detailed description of recent experimental 
 and theoretical 
 works on ferroelectricity in nanostructured materials and thin
 films, we refer the reader to recent book chapters
 \cite{Ghosez-06,Rabe-07,Lichtensteiger-07.2} and topical review
 papers~\cite{Dawber-05,Rabe-05,Ponomareva-05,Duan-06.2,Scott-06,Setter-06,
 Junquera-08,Zubko-11,Rondinelli-11}.

\subsection{Mechanical boundary conditions: strain} 
\label{s:strain}

 The coupling between ferroelectric polarization and strain is well 
 known to be especially strong in ferroelectric perovskite oxides,
 and can have a substantial impact on the structure,
 transition temperatures, dielectric and piezoelectric responses.
 Exhaustive discussions on strain effects in ferroelectric thin films
 can be found in Ref.~\cite{Schlom-07} (combined experimental and theoretical
 report), and in Refs.~\cite{Rabe-05} and \cite{Dieguez-08} (more focused
 on the theoretical point of view).

 In ferroelectric thin films, homogeneous biaxial strain can be achieved thanks to the epitaxial growth of the film on a substrate with a different lattice parameter. 
 Assuming perfect coherency, the strain is defined as a function 
 of the bulk lattice parameters of the film material, $a_{\rm f}$, and 
 of the substrate, $a_{\rm s}$, as $\eta = \frac{a_{\rm s}-a_{\rm f}}{a_{\rm f}}$. 
 The clamping between the film and the substrate onto which 
 it is deposited can be maintained only in ultrathin films,
 where the elastic energy stored in the overlayer is still 
 relatively small.
 For thicker films a progressive relaxation will occur
 via formation of misfit dislocations, which generally 
 cause a degradation in film quality.
 Note that, in general, the strain state of the film 
 will also depend on 
 the thermal evolution of the lattice parameters from the deposition 
 temperature to the measurement temperature, and the degree to which 
 full lattice coherence has been developed and maintained during film growth. 

 In a simplified theoretical model~\cite{Junquera-08}, 
 the main effect of the ``polarization-strain'' coupling
 is the renormalization of the quadratic term of the 
 free energy, like the one shown in Eq.~(\ref{eq:landau-expansion}).
 So, playing properly with the epitaxial strain conditions,
 it is possible to make the coefficient $\alpha_2$ more negative
 (further stabilizing the ferroelectric state or even inducing ferroelectricity in a non-ferroelectric material),
 or to make $\alpha_2$ positive, thus suppressing the ferroelectric 
 character of the film.
 
 The first milestone theoretical work on the influence of strain on the ferroelectric polarization was by Pertsev~\etal~\cite{Pertsev-98},
 who identified the correct ``mixed'' mechanical boundary conditions of the problem 
 (fixed in-plane strains, and vanishing out-of-plane stresses),
 and computed the corresponding Legendre transformation of the 
 standard elastic Gibbs function to produce the 
 correct phenomenological free-energy functional to be minimized.
 Then they introduced the concept, now known as ``Pertsev phase diagram'',
 of mapping the equilibrium structure as a function of temperature and 
 misfit strain, which has proven of enormous
 value to experimentalists seeking to interpret the
 behaviour of epitaxial thin films and heterostructures.

 These kind of diagrams were produced for the most standard
 perovskite oxides [BaTiO$_3$~\cite{Pertsev-98,Pertsev-99,Choi-04},
 PbTiO$_3$~\cite{Pertsev-98}, SrTiO$_3$~\cite{Pertsev-00,Pertsev-02,Haeni-04},
 and Pb(Zr$_{x}$\-Ti$_{1-x}$)O$_{3}$ (PZT)
 solid solution~\cite{Pertsev-03}].
 This approach generally yields very accurate results around the 
 temperature/strain regime in which the model parameters were 
 fitted (usually near the bulk ferroelectric transition).
 In distant strain-temperatures regimes the uncertainty tends to
 increase and, as reported for BaTiO$_3$, different sets of DGL
 parameters may provide qualitatively different phase 
 diagrams~\cite{Pertsev-98,Pertsev-99}.

 On the first-principles front, studies of 
 misfit strain effects in single-domain perovskite-oxide thin films
 have been successfully carried out for several materials, most notably 
 BaTiO$_{3}$~\cite{Dieguez-04}, PbTiO$_{3}$ and PbTiO$_{3}$/PbZrO$_{3}$ 
 superlattices~\cite{Bungaro-04},  and SrTiO$_{3}$ (including the 
 tunability of its dielectric response~\cite{Antons-05}).
 Full sequences of epitaxially-induced phase transitions and the 
 values of the corresponding critical strains for eight different 
 perovskites were reported in Ref.~\cite{Dieguez-05}.   
 Strain-induced ferroelectricity in an otherwise non-ferroelectric
 material was also theoretically predicted in simple rocksalt 
 binary oxides (BaO and EuO) and superlattices (BaO/SrO)~\cite{Bousquet-10}.

 From all these theoretical studies, a general trend emerged for perovskite oxides strained on a (001) 
 substrate~\cite{Dieguez-05}:
 sufficiently large epitaxial compressive strains tend to favor a ferroelectric $c$-phase
 with an out-of-plane polarization along the [001] direction; conversely,
 tensile strains usually lead to an $aa$-phase, with an in-plane $P$ oriented
 along the [110] direction.
 The behavior in the intermediate regime is material-dependent,
 but the general trend is that the polarization rotates continuously 
 from $aa$ to $c$ passing through the [111]-oriented $r$-phase. 
 In non-ferroelectric perovskites like SrTiO$_3$ and BaZrO$_{3}$ the intermediate regime is non-polar,
 while in PbTiO$_{3}$ the formation of mixed domains 
 of $c$ and $aa$ phases could be favorable. 
 
 From the experimental side, there have been impressive advances
 as well. Thanks to the availability of several perovskite substrates
 with a wide variety of in-plane lattice parameters, 
 we can now tune the ferroelectric and related properties in thin films 
 by using the homogeneous strain almost as a continuous knob, which
 led to the coinage of the term ``strain-engineering''.
 For instance, Haeni~\etal~\cite{Haeni-04} observed room-temperature ferroelectricity in SrTiO$_3$ with an in-plane component of P. The effect can be rationalized in terms of the +1\% tensile strain imposed by the DyScO$_3$ substrate, and is rather dramatic considering that SrTiO$_3$ is paraelectric in the bulk. Another example of strain-engineering was demonstrated by Choi~\etal~\cite{Choi-04}, with a large enhancement 
 of ferroelectricity induced in strained \bto\ thin films epitaxially grown on single-crystal 
 substrates of GdScO$_3$ and DyScO$_3$. The strain resulted in a 
 ferroelectric transition temperature nearly 500 K higher than the 
 bulk one and a remnant polarization at least 250\%\ higher than 
 bulk \bto\ single crystals. 
 Very recently, a spectacular strain
 effect was demonstrated both experimentally and theoretically~\cite{Zeches-09,
 Hatt-10}: multiferroic BiFeO$_3$ films undergo an isosymmetric phase transition
 to a tetragonal-like structure with a giant axial 
 ratio~\cite{Bea_bfo} when grown 
 on a highly compressive substrate such as LaAlO$_3$. Furthermore, 
 both phases appear to coexist~\cite{Zeches-09} in some conditions, with a 
 boundary that can be shifted upon application of an electric field. 
 This appears to be 
 by far the largest experimentally realized epitaxial strain to date, of
 the order of 4-5 \%; the existence of this new phase of 
 BiFeO$_3$ was also predicted 
 to be promising for enhancing the magnetoelectric response of this 
 material~\cite{Jorge_bfo}.

 Finally, in the last few years groundbreaking works have proposed 
 new mechanisms, based
 on the coupling of spin, optical phonons and strain to design 
 new multiferroic materials in which magnetic (electric) polarization
 can be induced by an applied electric (magnetic) field~\cite{Fennie-06}. 
 In some antiferromagnetic paralectric insulators, 
 there is a coupling between infrared-active phonons and magnetism.
 In this case, it might happen that the lowest frequency polar modes become
 unstable at a critical strain, producing an epitaxial-strain-induced 
 ferroelectricity in the antiferromagnetic state.
 The strain is not only coupled with the polarization but also
 with the magnetic ordering.
 If the ferromagnetic spin alignment softens more 
 a low-frequency polar mode that
 is strongly coupled to epitaxial strain, then the energy of the
 ferromagnetic-ferroelectric structure might drop below those of the 
 antiferromagnetic-ferroelectric phase for a given value of strain.
 This spin-phonon coupling mechanism for epitaxial-strain-induced 
 multiferroelectricity has been theoretically proposed for
 SrMnO$_{3}$~\cite{Lee-10.2} (critical tensile strain of +3.4 \%),
 and EuTiO$_{3}$~\cite{Fennie-06}, whose experimental demonstration
was made only recently~\cite{Lee-10.1} (critical biaxial strains
 of -1.2 \% under compression and +0.75 \% under expansion).

As mentioned above, coherency between the substrate and
the thin films lattice constants can be maintained
only up to a limiting thickness,
after which defects and misfit dislocations start to form. Strain relaxation leads to \emph{inhomogeneous} strain fields (or strain gradients), which can have profound consequences on the properties of the thin film. A strain gradient intrinsically breaks the spatial inversion symmetry and hence acts as an effective field, generating electrical polarization even in centrosymmetric materials. This phenomenon became known as \emph{flexoelectricity}, by analogy with a similar effect in liquid crystals, and is alowed in materials of any symmetry. Strain-gradient-induced polarization has, for instance, been measured in single crystals of SrTiO$_3$, a non-polar material~\cite{Zubko-07}.

Flexoelectric effects can play an important role in the degradation of ferroelectric properties~\cite{Catalan-04,Catalan-05} and therefore proper management of strain gradients is crucial to the perfomance of ferroelectric devices. At the same time, an increasing amount of research is now aimed at exploiting flexoelectricity for novel electronic devices. The possibility of generating a flexoelectric response in any dielectric material ~\cite{Kogan-64}, irrespective of its symmetry, by carefully engineering strain gradients has generated a lot of excitement in the field (see, for instance, the review by L. E. Cross~\cite{Cross-06}). At the same time, fundamental questions about flexoelectricity are not completely settled, and modern first-principles-based approaches ~\cite{Resta-10.2, Hong-10} are being developed to revisit existing phenomenological theories from a microscopic perspective~\cite{Tagantsev-86,Tagantsev-91}. 

\subsection{Electrical boundary conditions: imperfect screening}
\label{s:ebc}

\begin{figure}[!htb] 
    \begin{center}
       \includegraphics[width=10cm,keepaspectratio]{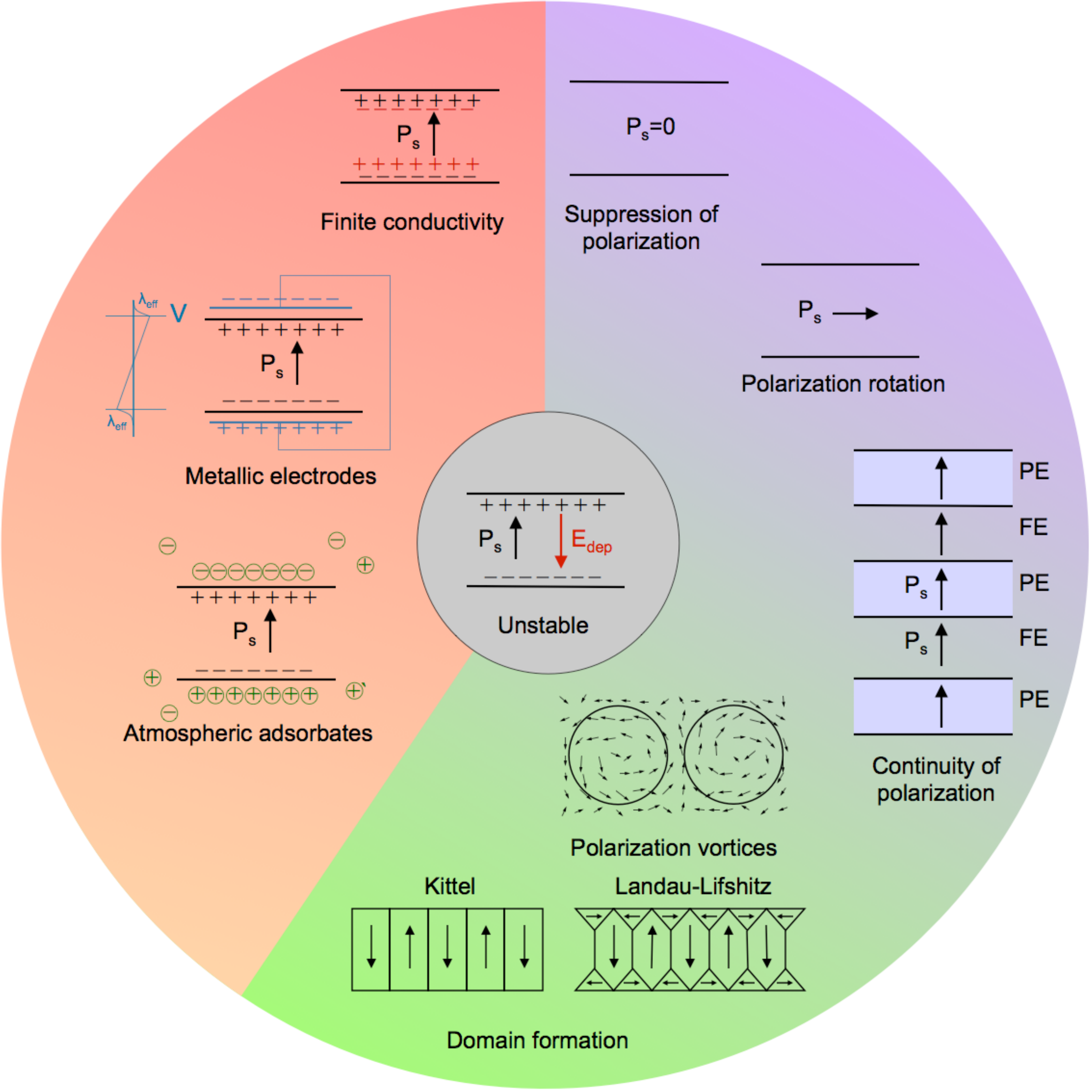}
       \caption{ The depolarization field arising from unscreened bound 
                 charges on the surface of the ferroelectric is generally 
                 strong enough to suppress the polarization completely 
                 and hence must be reduced in one of a number of ways if 
                 the polar state is to be preserved. Much of the research 
                 on ultrathin ferroelectrics thus deals directly or 
                 indirectly with the question of how to manage the 
                 depolarization fields. The left part of the diagram 
                 illustrates screening by free charges from metallic electrodes,
                 atmospheric adsorbates or mobile charges from within 
                 the semiconducting ferroelectric itself. Note that even 
                 in structurally perfect metallic electrodes, the 
                 screening charges will spread over a small but finite length, 
                 giving rise to a non-zero effective screening length 
                 $\lambda_{\rm eff}$ that will dramatically alter the 
                 properties of an ultrathin film. Even in the absence of 
                 sufficient free charges, however, the ferroelectric 
                 has several ways of preserving its polar state, 
                 as shown in the right part of the diagram. One possibility 
                 is to form domains of opposite polarization that lead to 
                 overall charge neutrality on the surfaces (Kittel domains) 
                 or closure domains (Landau-Lifshitz domains). 
                 Under suitable mechanical boundary conditions, 
                 another alternative is to rotate the polarization 
                 into the plane of the thin ferroelectric slab. 
                 In nanoscale ferroelectrics polarization rotation 
                 can lead to vortex-like states generating ``toroidal'' order. 
                 In heterostructures such as ferroelectric(FE)-paraelectric(PE) 
                 superlattices, the non-ferroelectric layers may polarize 
                 in order to preserve the uniform polarization state 
                 and hence eliminate the depolarization fields. 
                 If all else fails, the ferroelectric polarization 
                 will be suppressed.
               }
       \label{f:Screening}
    \end{center}
 \end{figure}

The depolarization field arising from unscreened bound 
                 charges on the surface of a ferroelectric thin film is generally 
                 strong enough to suppress the polarization completely 
                 and hence must be reduced in one of a number of ways if 
                 the polar state is to be preserved (see Fig.\ref{f:Screening}). Much of the research 
                 on ultrathin ferroelectrics thus deals directly or 
                 indirectly with the question of how to manage the 
                 depolarization fields. The screening can be obtained by free charges from metallic electrodes,
                 ions from the atmosphere or mobile charges from within 
                 the semiconducting ferroelectric itself. Note that even 
                 in structurally perfect metallic electrodes, the 
                 screening charges will spread over a small but finite length, 
                 giving rise to a non-zero effective screening length 
                 $\lambda_{\rm eff}$ that will dramatically alter the 
                 properties of an ultrathin film. The problem of imperfect screening and how to model it will be addressed here. Later, we will see that even in the absence of 
                 sufficient free charges, the ferroelectric 
                 has several other ways of preserving its polar state, e.g., by forming domains of opposite polarization, or trotating the polarization
                 into the plane of the thin ferroelectric slab.

 It is well known than in most experiments on thin-film perovskite
 capacitors the electrical properties of the device tend to be
 plagued by deleterious size effects.
 In the paraelectric regime (i.e. above the Curie 
 temperature) the measured capacitance tends to be orders of
 magnitude smaller than expected from the classical formula
 $C=\epsilon S/t$, where $\epsilon$ is the dielectric constant of the 
 insulator, $t$ is the thickness of the dielectric layer, and $S$ the 
 surface area of the capacitor plates.
 In the ferroelectric regime the 
 remnant polarization decreases with the thicknesses of the
 film~\cite{Nagarajan-04,Lichtensteiger-05,Jo-05,DJKim-05,Helvoort-05}. Eventually, the film loses its ferroelectric properties below a certain \emph{critical thickness}~\cite{Junquera-03.1,Junquera-08}, or alternatively breaks into domains of opposite polarization~\cite{Nagarajan-06}
 (we shall present a more extensive discussion of domains in 
 Sec.~\ref{s_domains}).
 The Curie temperature, $T_{\rm c}$, itself tends to
 shift to lower values, and the dielectric anomaly (which 
 in bulk is a sharp peak, directly correlated to the ferroelectric 
 transition) is smeared out so that interpretation of the 
 measurements is sometimes difficult~\cite{Vendik-00,Lookman-04}. 
 Fast relaxation of the polarization back to an unpolarized 
 state after switching off the poling bias was  
 reported~\cite{Kim_et_al:2005}; the coercive field
 itself is known to be significantly larger compared to the 
 bulk value.

 Traditionally, the origin of these size effects
 was attributed to the existence of a physically and chemically 
 distinct passive layer with degraded ferroelectric properties
 (and/or with smaller permittivity than the film material)
 at the ferroelectric/electrode interface, the so-called ``\emph{dead layer}''.
 Many processing issues can lead to the formation of such a passive 
 layer. For example, damage can be created at the film surface
 by bombardment of the sputtering ions during the deposition of
 a top electrode~\cite{Larsen-94}, or by changes in stoichiometry due to
 the loss of volatile elements such as lead or oxygen.
 A non-switching dead layer~\cite{Larsen-94,Miller-90,Tagantsev-95}
 can also derive from pinning of domain-walls~\cite{Lebedev-94}, or
 the screening of the internal field by a depletion
 region~\cite{Tagantsev-97}. 
 The formation of oxygen vacancies during growth can be another 
 important factor to be cautious about, as they might lead
 to the presence of space charge in the film; the detrimental 
 consequences of space charge near the surface have been known
 since the work of K\"{a}nzig in 1955~\cite{Kanzig-55}.
 Note that in general oxygen vacancies seem to be implicated in 
 most of the failure mechanisms of ferroelectric capacitors,
 e.g. their redistribution with electrical cycling is believed to
 be the main cause of polarization fatigue~\cite{Yoo-92,Dawber-00,Lo-02}. 

 While the above arguments have traditionally supported an extrinsic 
 origin of the dead layer, more recently increasing attention has been paid to possible intrinsic effects. 
 Indeed, the improvement of growth techniques and the widespread use 
 of metallic perovskites as electrode materials have allowed for the 
 realization of capacitor structures with nearly ideal lattice-matched 
 interfaces.
 In these high-quality films the aforementioned processing concerns 
 were minimized, but the size effects turned out to still be significant. 
 This suggests that even an ideal ferroelectric/metal interface might 
 be characterized by an effective ``dead layer''; this would stem from
 the fundamental quantum-mechanical properties of the junction rather
 than from growth-induced defects.
 Before the advent of first-principles simulations for this class of 
 systems, a popular semiclassical model invoked the finite Thomas-Fermi 
 screening length, $\lambda_{\rm TF}$, of realistic electrodes (as 
 opposed to an ideal metal, which would perfectly cancel the surface 
 polarization charges of the ferroelectric)~\cite{Black/Welser-99}.
 This would create a spatial separation between bound charges and
 screening free charges, i.e. it would act as a capacitance in 
 series with the film, with detrimental effects that are in all
 respects analogous to those of a physically degraded layer near the 
 interface.
 In fact, the idea of an imperfect screening has considerable history; 
 in the 70s~\cite{Mehta-73,Batra-72,Batra-73,Wurfel-73}, 
 within the framework of the Landau theory,
 researchers at IBM found that a finite $\lambda_{\rm TF}$ would lead to the
 appearance of a residual depolarization field, change 
 the order of the phase transition, 
 reduce the magnitude of the polarization and shift the 
 transition temperature.

 At the \emph{microscopic} level the reliability of these semiclassical 
 arguments is, however, unclear. Contrary
 to the case of a doped semiconductor, $\lambda_{\rm TF}$
 in typical electrode materials is fairly small, usually 0.5 \AA{} or less.
 This implies that the Thomas-Fermi screening would occur in a region that lies
 adjacent to the interface plane, where the properties of the electrode 
 material are far from bulk-like. The use of bulk electrode parameters
 in such a context (e.g. $\lambda_{\rm TF}$) appears therefore unjustified.
 Furthermore, the quantum-mechanical penetration of conduction states into
 the insulating film might (at least partially) cancel the effects of a finite
 $\lambda_{\rm TF}$; this is usually neglected in semiclassical models.
 Finally, interface-specific effects and chemical bonding may be as important as purely electronic
 effects; again, these effects are absent from the simplified Thomas-Fermi 
 treatments.

 First-principles theory have played a leading role in recent years 
 at addressing the above questions with unprecedented accuracy, as in a
 typical density-functional calculation all the electrostatic and chemical 
 effects occurring at a realistic interface are automatically included at 
 a fundamental, unbiased level.
 The seminal works of Ref.~\cite{Junquera-03.1} and Ref.~\cite{Stengel-06}
 have demonstrated, with fully quantum-mechanical calculations, the 
 existence of an interface-related depolarization effect in ferroelectric
 and paraelectric capacitors, respectively. 
 Many works have followed during the past few years, pointing out
 important aspects of the problem, such as the ionic contribution to
 the screening (predicted in Refs.~\cite{Sai-05,Gerra-06} and 
 experimentally observed by high-resolution transmission electron
 microscopy in Refs.~\cite{Jia-07,Chisholm-10}), or the role of
 the electrode material in aggravating or reducing the size
 effects~\cite{Sai-05,Stengel-06}.
 The findings of these investigations can be roughly 
 summarized into three main messages:
  (i) an ``intrinsic dead layer'' is, in general, present even at 
  perfect, defect-free interfaces~\cite{Stengel-06}; (ii)
  the magnitude of the depolarization effect is an \emph{interface}
  property (i.e. it depends on materials, crystallographic orientation
  and lattice termination), and is best described as an \emph{effective
  screening length} $\lambda_{\rm eff}$ (or, equivalently, interface 
  capacitance) that includes all the electronic and ionic effects 
  described above; (iii) that use of simple-metal electrodes (such as Pt 
  or Au) is almost systematically predicted to be beneficial for the electrical 
  properties of the device as compared to oxide electrodes.
 Point (iii) above is particularly puzzling in light of the experimental
 results, which show a much better performance when oxide electrodes
 (e.g. SrRuO$_3$) are used. The use of SrRuO$_3$ was indeed
 an experimental breakthrough in that it allowed for a drastic reduction 
 of fatigue issues~\cite{Ramesh-91,Ramesh-92}.
 It is important to stress that the first-principles results are 
 relevant for an ideal, defect-free interface; this is, unfortunately, 
 very challenging to realize in practice in the case of Pt electrodes.
 In addition to the practical question ``Which electrode is better?'',
 first-principles theory is able to answer more fundamental (and 
 maybe more interesting) ones, such as: ``Why are ideal simple-metal 
 electrode/ferroelectric interfaces behaving better?'' 
 We shall come back to this point in Sec.~\ref{s:chemical}; in the
 following Section we first briefly discuss the consequences of a finite
 $\lambda_{\rm eff}$ on the electrical properties of the device.

 Before closing this Section on the microscopics of the 
 polarization screening mechanism, it is useful to mention an
 increasingly large body of literature that concerns ferroelectric 
 films with an open surface, i.e. without a top electrode. 
 In this case, as there are no metallic free carriers
 available, adsorbates or point 
 defects~\cite{Watanabe-01,Spanier-06,Fong-06,Kolpak-07,Li-08}
 are believed to supply the compensating surface charges that
 are necessary to stabilize a polar state.
 Bringing these ideas one notch further, Wang {\em et al.} have
 recently demonstrated reversible and reproducible ferroelectric
 switching in a thin PbTiO$_3$ film by varying the partial oxygen 
 pressure pO$_{2}$ at the open surface~\cite{RVWang-09}.
 This breakthrough proof of concept was coined ``chemical switching''.

\subsection{Electrical functionals with a depolarization field}
\label{s:efdp}

 Whatever the microscopic origin of the depolarization effects, a typical 
 parallel plate capacitor with non-ideal electrodes can be modeled in terms
 of a bulk-like film that is $N$ unit cells thick, 
 in series with two interfacial
 layers, of thickness $\lambda_{\rm I}$, that behave like a linear dielectric
 material of dielectric constant $\epsilon_{\rm I}$.
 In the following, we assume an uniaxial ferroelectric and 
 keep only one component
 of all the vector quantities.
 Then, the energy as a function of the electric displacement $D$ can be written
 as the energy of the three capacitors in series: the two (assumed identical)
 interfaces and the bulk of the film ($D$ can be viewed as the surface density of free 
 charges stored on the plates)~\cite{Stengel-11}:
 \begin{equation}
    U_N(D) = N U_{\rm b}(D) + 2S \lambda_{\rm eff} \frac{D^2}{2\epsilon_0}.
    \label{eq:ucapa}
 \end{equation}
 Here $U_{\rm b}$ is the bulk internal energy per unit cell at
 a given $D$, $S$ is the cell cross-section, $\epsilon_{0}$ is the free vacuum permittivity, and 
 $\lambda_{\rm eff}=\lambda_{\rm I}/\epsilon_{\rm I}$ is the
 effective screening length mentioned in the previous Section (only the
 ratio between $\lambda_{\rm I}$ and $\epsilon_{\rm I}$ is physically
 relevant).
 The equilibrium monodomain state of such a capacitor in short-circuit
 is readily obtained by imposing $dU_N(D)/dD$=0, which leads to the
 condition $N dU_{\rm b}/dD = - 2S \lambda_{\rm eff} D / \epsilon_0$.
 Note that the internal (``depolarization'') field in the bulk-like region 
 is \emph{defined} as $\Omega \bm{E}_{\rm d} = dU_{\rm b}/dD$, 
 where $\Omega=Sc$ is the unit cell volume and $c$ is the out-of-plane 
 lattice parameter.
 Note also that in typical ferroelectrics the electric displacement can 
 be very accurately approximated with $P_0$, the ferroelectric polarization 
 associated to the soft mode; then, by writing the total thickness $t=Nc$ 
 we obtain the depolarization field,
 \begin{equation}
    \bm{E}_{d} = - \frac{2 \lambda_{\rm eff} P_0}{ \epsilon_{0} t},
    \label{eq:depolfield}
 \end{equation}
 From Eq. (\ref{eq:depolfield}) we see that the depolarization
 field depends linearly on the effective screening length and
 the polarization, $P_0$, and is inversely proportional to the film 
 thickness $t$.
 The minus sign means that the field opposes to the polarization.
 The field has to remain small enough in order to stabilize
 a monodomain polar state in the capacitor.
 All the quantum effects and the chemistry at the interface 
 are embedded in the only parameter $\lambda_{\rm eff}$,
 that will depend on the ferroelectric material, the electrode, the atomic
 structure and the particular orientation of the interface.
 An alternative derivation of Eq.~(\ref{eq:depolfield}) can be found in 
 Ref.~\cite{Dawber-03,Ghosez-06,Junquera-08}, and a 
 detailed comparison between the expressions given by different models is 
 available in Ref.~\cite{Ghosez-06}.

 As it is clear from previous discussions, the electrical boundary conditions
 are extremely important in the stabilization of the different phases
 and on the functional properties of the capacitors.
 It also appears clear that layered geometries such as capacitors or 
 monodomain superlattices can be very effectively described within a 
 simple series capacitor model, where $D$ is the fundamental electrical
 variable, while using $E$ would impose short-circuit boundary conditions across the whole simulation box due to the overall periodicity of the system.
 In this context, the recently-developed first-principles techniques 
 (discussed earlier in the Chapter) to treat $D$ as the fundamental electrical
 variable~\cite{Stengel-09-np, Stengel-09.2} appear ideally suited to 
 perform this kind of decomposition.

 The preservation of $D$ across the heterostructure leads a ``locality 
 principle'', where the local bonding effects can be unambiguously
 separated from the long-range electrostatics interactions, allowing
 for a detailed analysis of the microscopic mechanisms contributing to the
 polarization.
 This opens the door to LEGO-like models, 
 where the electrical properties of \emph{individual} atomic 
 layers~\cite{Wu-08}, even including oxygen octahedra rotations~\cite{Wu-11}, 
 are computed and stored as a function of the in-plane strain and electric 
 displacement, and combined afterwards in arbitrary manners to predict 
 polarization or nonlinear dielectric response of user-designed superlattices.

\subsection{Chemical-bonding contributions to the electrical boun\-dary conditions}
\label{s:chemical}

 We have seen in the previous sections that, in standard ferroelectric
 capacitors, the electrode/ferroelectric junction generally introduces
 an additional term in the electrostatic energy functional of the film.
 First-principles simulations demonstrated that such a term, which 
 typically tends to suppress ferroelectricity via a depolarization field 
 in zero bias, is present even in the case of a perfect defect-free 
 interface, where it was shown to depend on the microscopic details of 
 the junction.
 What we have not discussed yet is the precise relationship between 
 these microscopic properties (e.g. local electronic and ionic structure, 
 chemical bonding, etc.) and the macroscopically relevant quantities,
 e.g. $\lambda_{\rm eff}$.
 Such an analysis was performed in two recent first-principles 
 works,~\cite{Stengel-09-nm,Stengel-09.2} where the techniques to
 work at constant $D$ were used to achieve a local decomposition 
 between bulk and interface contributions to the electrical 
 equation of state of a thin-film capacitor.
 Based on this decomposition, the interface-specific ingredients could
 be singled out, as we shall explain in the following.

 First, recall that the traditional phenomenological understanding of
 the imperfect screening at a metal/ferroelectric interface was based on 
 a Thomas-Fermi model of the free carriers in a real metal, i.e.
 on purely \emph{electronic} effects.
 It is not difficult to realize that this gives only a partial picture,
 and that chemical bonding effects should be taken into account as well
 -- possibly even more seriously than electronic ones.
 Why? Because the high dielectric constant $\epsilon$ of perovskite ferroelectrics
 is largely dominated by lattice polarization. The ionic lattice mediates
 a contribution of the type $\Delta \epsilon \sim (\bar{Z}^*)^2 / \omega^2$,
 where $\omega$ and $\bar{Z}^*$ are the frequency and the Born effective 
 charge of a given zone-center optical mode. As the soft-mode frequency goes 
 to zero, $\Delta \epsilon$ tends to diverge.
 Now recall that the vanishing frequency of the soft mode results from a subtle balance between
 long-range and short-range forces. If at the interface we create strong 
 chemical bonds between the film and the electrode, we inevitably stiffen the
 short-range part, and this might raise $\omega$ considerably. This would
 create an interface oxide layer with lower local $\epsilon$, i.e. a dielectric
 ``dead layer''.
 Conversely, if one carefully engineers the interface in such a way that the
 metal/oxide bonds are loose, then the interface-related suppression of the
 local dielectric constant should be minimized, or even completely avoided.
 In principle, this picture suggests an even more tantalizing possibility,
 that is, of making the interface bonds loose enough that the surface unit
 cell becomes ferroelectrically active, possibly even more so than the 
 bulk film material itself. 
 This would constitute a drastic departure from the conclusions of 
 the Thomas-Fermi theory, which systematically predicts a deleterious
 depolarization effect at a metal/ferroelectric interface.
 
 As unrealistic as it may sound, such a scenario was recently demonstrated 
 by means of first-principles calculations in Refs.~\cite{Stengel-09-nm,Stengel-09.2}.
 These works focus on BaO-terminated BaTiO$_{3}$ thin films that are
 symetrically sandwiched between two Pt electrodes. 
 At this interface the Ba and O atoms that terminate the BaTiO$_{3}$ 
 lattice sit atop the square lattice of Pt atoms at the 
 (100)-oriented electrode surface.
 Pt chemically binds to O, while it repels Ba. This competition 
 stretches and weakens the Pt-O bond, that becomes ``ferroelectrically
 active''. This mechanism produces an enhancement of the overall
 ferroelectricity of the capacitor, which becomes stronger the 
 thinner the film.
 At the extreme limit of a two-unit cell BaTiO$_3$ film the 
 spontaneous (and \emph{switchable}) polarization is 35 \% 
 larger than the bulk BaTiO$_{3}$ value.~\cite{Stengel-09-nm}
 This implies an opposite behavior to the traditional ``dead layer'' 
 effect, which could be rationalized within the electrostatic model
 of the previous section by assuming a \emph{negative} $\lambda_{\rm eff}$.
 The AO-terminated perovskite/simple metal interface is of course 
 a very specific interface structure, that might or might not be 
 representative of the experimentally realizable stable configurations. 
 It was mainly chosen as a proof of principle, i.e. as a testcase 
 where the ideas sketched in the above paragraphs happen to 
 manifest themselves in a particularly dramatic way. It should be
 kept in mind that the above principles about the relationship between
 bonding and dielectric properties are completely general, and we
 expect them to play an important role at \emph{any} metal/ferroelectric
 interface. We stress that the macroscopic electrical parameter 
 $\lambda_{\rm eff}$ does include implicitly all the effects of 
 chemical bonding, together with purely electronic contributions. 
 In that sense, it is not entirely appropriate to speak of ``chemical 
 boundary conditions'' as something distinct from the electrical boundary
 conditions: the former are best understood as chemical-bonding 
 contributions to the latter.

\section{Polarization domains and domain walls}
\label{s_domains}

The previous section was concerned predominantly with ultrathin ferroelectrics sandwiched between metallic electrodes that provide the  screening charges necessary to  stabilize the ferroelectric ground state. Even in the absence of free charges, however, ultrathin ferroelectrics find a number of ways of preserving their polar state, as illustrated in Fig.~\ref{f:Screening}. One possibility is to form domains of opposite polarization known as Kittel domains, or flux closure (Landau-Lifshitz) domains.

\subsection{Kittel law}
\label{s_Kittellaw}

Domain formation in ferroelectrics is analogous to that of ferromagnets and thus most of the theory was adapted from the seminal works of Landau and Lifshitz \cite{Landau-35} and Kittel \cite{Kittel-46,Kittel-49}.  Domains lead to overall charge neutrality at the surfaces,  eliminating the depolarization field, at least throughout the majority of the ferroelectric slab, and  reducing the electrostatic energy of the system. For Kittel domains (Fig.~\ref{f:Screening}), the stray fields arising from the antiparallel arrangement of dipoles, are confined to a thin near surface region and decay exponentially over a lengthscale comparable to the domain width $w$. The  electrostatic energy of these fields thus increases with domain size as          
         \begin{equation}
            F_{P}\sim P_{\rm s}^2 w,
         \end{equation}
favoring  domains that are as small as possible. For Landau-Lifshitz domains (Fig.~\ref{f:Screening}), a similar term linear in $w$ exists but has its origin in the anisotropy energy---the cost of rotating the dipoles near the surface.

Decreasing the domain size, however, also increases the density of domain walls---the boundaries that separate domains of different polarization. Short-range dipolar interactions resulting from the modification of the dipole moments when passing through a domain wall give rise to an energy cost know as domain wall energy $\sigma_W$ per unit area of the wall, and contribute a term that scales with domain wall density, i.e. as $1/w$ to the total energy
         \begin{equation}
            F_{W} = \sigma_{W} \frac{t}{w},
         \end{equation}
where $t$ is the thickness of the ferroelectric. Minimizing $F_W+F_P$ with respect to $w$ gives the famous Kittel relation 
\begin{equation}
   w \sim \sqrt{t}.
\label{eq:Kittel}
\end{equation}

\begin{figure}[!htb] 
   \begin{center}
      \includegraphics[width=8cm,keepaspectratio]{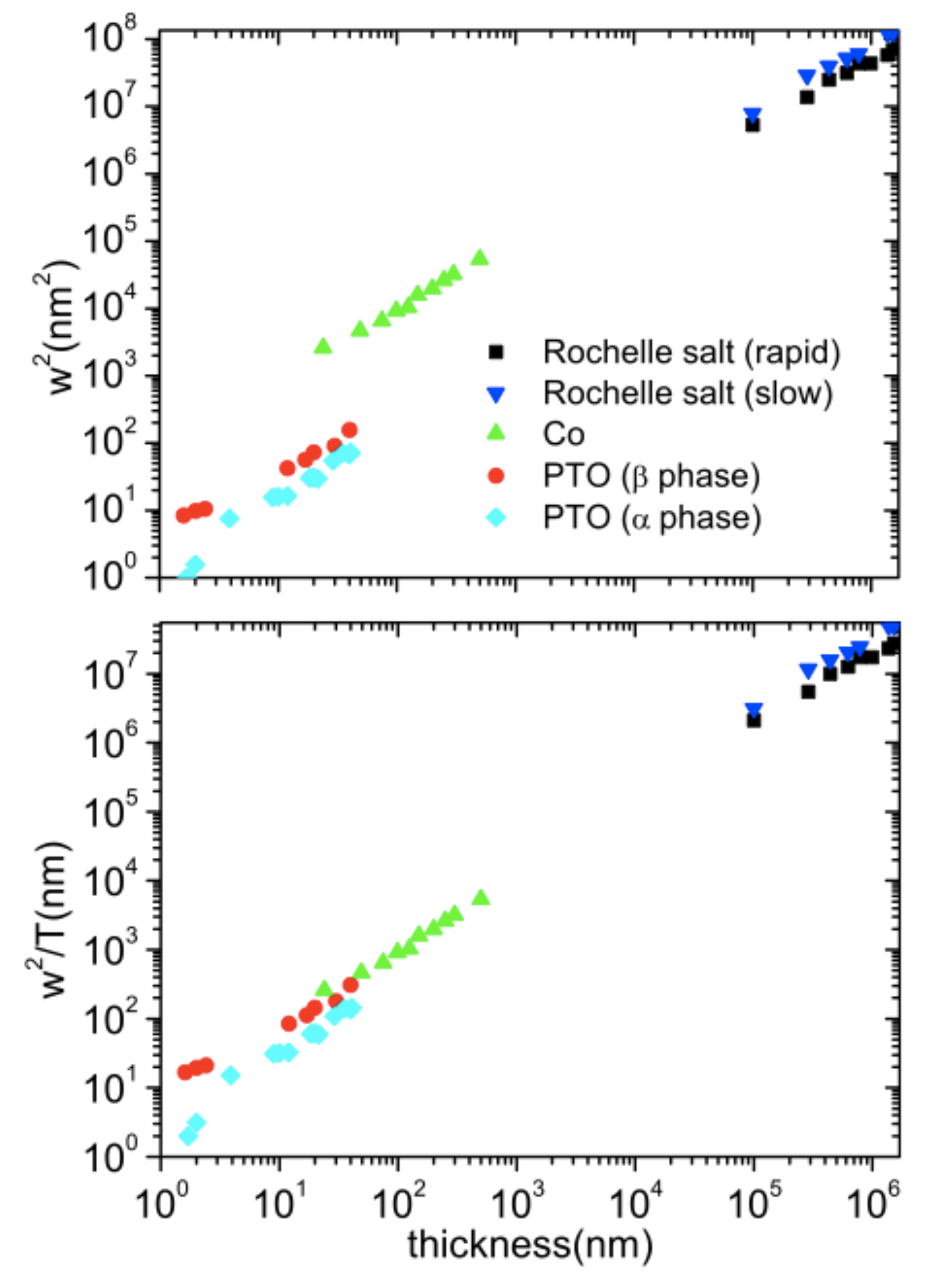}
      \caption{(Above) Square of the 
               180$^\circ$ domain width as a function of crystal 
               thickness for some ferroics (data extracted from 
               Ref.~\cite{Mitsui-53,Streiffer-02,Hehn-96}).
               (Below) When the square of the domain size is divided by 
               the wall thickness, all data fall on the same parent curve.
               The wall thicknesses used for the scaling have been
               extracted from 
               Ref.~\cite{Mitsui-53,Hubert-98,Meyer-02}.
               Reprinted with permission from
               Ref.~\cite{Catalan-07.2}. 
               }
      \label{f:CatalanPowerLaw}
   \end{center}
\end{figure}

As device dimensions shrink, so do the domains. Remarkably,  the validity of this simple scaling law for ferroelectric materials extends over six orders of magnitude in $t$, as shown in  Fig.~\ref{f:CatalanPowerLaw} \cite{Catalan-07.2}. The constant of proportionality in (\ref{eq:Kittel}) depends on a number of material properties, generally yielding much larger domain widths for ferromagnets than for ferroelectrics\cite{Schilling-06}. If the domain wall thickness $T$ is taken into account, however, a universal dimensionless quantity $w^2/Tt$ can be defined for all ferroic domains  \cite{Scott-06}, see Fig~\ref{f:CatalanPowerLaw}.

Theoretically, the Kittel law has been demonstrated within the 
model Hamiltonian approach for Pb(Zr$_{0.4}$Ti$_{0.6}$)O$_{3}$ 
thin films~\cite{Lai-07.2}.
Perfect scaling according to this law was obtained for thicknesses above 
4 unit cells.  In PbTiO$_3$, Streiffer~\etal~\cite{Streiffer-02,Fong-04} 
observed by in-situ x-ray synchrotron diffraction the evidence of 180$^\circ$ 
stripe domains with periods of 1.2-10 nm (going as $\sim \sqrt{t}$) 
in $c$-axis oriented thin films from 420 down to 12 \AA\ thickness 
grown epitaxially and coherently on SrTiO$_3$ insulating substrates.

It is worth noting that  Kittel's law can also be generalized to more complex three-dimensional shapes \cite{Catalan-07.1}.

 \subsection{Domains morphology}
\label{s_domainmorphology}

For a long time it was believed that domains in ferroelectrics are of the Kittel type, due to the strong crystalline anisotropy that makes polarization rotation difficult. Recent theoretical work, however, has shown that significant polarization rotation is expected in many cases, particularly as dimensions are reduced, fuelling an intense experimental search for evidence of this. Building on earlier phenomenological theories \cite{Kopal-97,Bjorkstam-67,Bratkovsky-00,Pompe-93,Pertsev-95,Bratkovsky-01.2,Stephenson-06},  model Hamiltonian and \emph{ab-initio} calculations have revealed that the precise domain morphology is expected to depend strongly  on the electrical and mechanical boundary conditions, and will in general be of a character that is intermediate between that  of the  Kittel and Landau-Lifshitz models \cite{Prosandeev-07, Lai-07.2, Aguado-Puente-08, Shimada-10}. Static and dynamic properties of such stripe domains have been calculated for BaTiO$_3$ \cite{Tinte-01}, PbTiO$_3$,  Pb(Zr,Ti)O$_3$ \cite{Kornev-04,Lai-06,Ponomareva-06,Wu-04a} and BiFeO$_3$ \cite{Prosandeev-10}, as well as superlattices consisting of periodically alternating layers of BaTiO$_3$ and SrTiO$_3$ \cite{Lisenkov-09}, giving important insight into the dependence of domain morphology on the precise crystalline and chemical structure of the material and revealing other types of domain patterns, such as \emph{``bubble nanodomains''} \cite{Kornev-04}. The real picture is in addition complicated by the presence of disorder that affects both the static morphology and dynamic behavior of nanodomains \cite{Paruch-07, Catalan-08}. Experimental findings and theoretical understanding of ferroelectric (and ferroelastic) domains in bulk crystals and thin films developed during the past 60 years can be found in Ref.~\cite{Tagantsev-2010}.

Reducing the dimension not just in one direction as is the case 
for ultrathin films, but in two (nano-rods~\cite{Urban-02}, 
-wires~\cite{Yun-02}, -tubes~\cite{Luo-03}) or even the three directions 
in space (nano-dots~\cite{OBrien-01,Schilling-09} 
and -particles~\cite{Liu-01}) leads to more complex domain patterns.  Theoretical studies of ferroelectric nanoscale disks and rods \cite{Fu-03,Naumov-04,
Ponomareva-05.2,Prosandeev-06.1,Prosandeev-07.2,Prosandeev-08.1,Prosandeev-09} 
have revealed yet another mechanism for dealing with the depolarization field:  formation of polarization vortices with zero net polarization, but a finite
spontaneous toroid moment. 
This ``ferroelectric'' state does not seem to be affected by the 
surface local environments, unlike that of an ultrathin 
ferroelectric film. Interestingly, the toroid moment can  be equivalently 
parallel or anti-parallel to the $z$-axis, and it is possible to switch 
from one minimum of the toroidal phase to the other by applying a 
time-dependent magnetic field (the magnetic field interacts with the 
total toroid moment of the nanoparticles by generating a curling 
electric field $\nabla\times E=-\partial B/\partial t$), inciting some researchers to dream about novel bistable devices based on toroidal order 
Ref.~\cite{Prosandeev-08.2}. It is to be noted however that, 
although direct observations of vortex states in soft ferromagnetic 
nanodots and nanorings~\cite{Shinjo-00,Wachowiak-02,Choe-04,Zhu-06}
have been made, experimental evidence in ferroelectrics is much more scarce.
A few recent experiments tend to confirm the existence of such  
toroidal polarization ordering, as for example in Pb(Zr,Ti)O$_3$ nanodots 
observed by piezoresponse force microscopy~\cite{Rodriguez-09}. Recent developments in sphetical aberration correction for transmission electron microscopes have been invaluable in revealing polarization rotation for the first time with atomic resolution~\cite{Nelson-11, Jia-11}. Mesoscopic dipole closure patterns were revealed by piezoresponse force microscopy images on free-standing, single-crystal lamellae of BaTiO$_3$~\cite{McQuaid-2011}.

\subsection{Domain walls}

Domain walls –- the boundaries between domains -– are interesting in their own. New experimental and theoretical work has revealed that they often possess fascinating structural and electronic properties that are absent in the parent material, opening an exciting possibility for novel devices based on the concept of domain wall engineering~\cite{Salje-2009}. An interesting example is the observation of a sizeable spontaneous polarization in (100) twin walls in CaTiO$_3$, a ferroelastic paraelectric material~\cite{Goncalves-Ferreira-2008}. Possible polar ferroelastic domain walls have also been found in paraelectric SrTiO$_3$~\cite{Zubko-07}. Other remarkable phenomena have been observed, e.g., twin walls that can support superconducting currents within WO$_{3-x}$, an otherwise insulating material~\cite{Aird-1998}. Our discussion here will focus on the domain walls in ferroelectric materials, and on the recent discovery that domain walls in multiferroic BiFeO$_3$ are conducting and can be used to obtain above-band-gap photovoltaic responses~\cite{Seidel-09,Yang-10}.

 BiFeO$_3$ is a largely studied multiferroic material, since at 
 room temperature it is ferroelectric as well as anti-ferromagnetic. 
 The polarization in BiFeO$_3$ is directed toward one of the 
 8 equivalent $<$111$>$ directions in the pseudocubic unit cell. 
 Combining these different polarizations into pairs of neighbors, 
 it can be easily checked that there will be 3 different angles possible 
 between two different polarizations: 180$^\circ$, 109$^\circ$ and 71$^\circ$. 
 These define the three different types of domain walls that can be observed 
 in BiFeO$_3$ at room temperature. 

 Using room temperature conductive atomic force microscopy, 
 Seidel~\etal~\cite{Seidel-09} found surprisingly that 180$^\circ$ 
 and 109$^\circ$ domain walls in BiFeO$_3$ are conducting, 
 while 71$^\circ$ are not. To understand this observation, one might be naively tempted to invoke a ``polar discontinuity'' scenario, where a (hypothetical) orientation-dependent discontinuity in $P$ would induce a two-dimensional gas of conducting carriers. This is, however, not the case here: as a matter of fact, all these domain walls are charge-neutral, i.e. the normal component of the macroscopic $P$ is  preserved across the wall. Interestingly, high-resolution transmission electron 
 microscopy images of a 109$^\circ$ domain wall reveal a small \emph{local} increase in the Fe displacements (relative to the Bi lattice) perpendicular to the domain wall, within the same region where the parallel displacements invert their sign. An analogous displacement pattern was found in model DFT calculations of the same system, which further predict a step in the electrostatic potential of up to 0.18 eV. A potential step at BiFeO$_3$ walls was later found to be involved in carrier photogeneration processes~\cite{Yang-10}, and it is therefore not unreasonable to think that it might also be responsible for the enhanced conductivity reported by Seidel et al. 
 
To understand the origin of such an electrostatic potential step, it is useful to consider the simpler case of PbTiO$_3$, where the spontaneous polarization is oriented along (100). In the case of 180$^\circ$ domain wall in PbTiO$_3$, with a net polarization on either side, the metal cations act as a center of inversion symmetry; therefore, a dipole moment or a potential step are both forbidden. Conversely, in the case of 90$^\circ$ walls in PbTiO$_3$, the two sides of the domain wall cannot be related by any symmetry operation; a net dipole moment is allowed, and calculations indeed confirm the presence of an associated potential step~\cite{Meyer-02}. Similar ideas would apply to BiFeO$_3$, although the situation here is slightly more complicated because of the (111) orientation of the spontaneous polarization. 

As an alternative explanation for the enhanced conductivity, Seidel et al. also observed that at the wall the fundamental band gap of BiFeO$_3$ might be reduced with respect to that of a uniform domain. This hypothesis is again supported by the DFT data. A decrease in the band gap would modify the band offset with a conductive atomic force microscopy tip and possibly favor carrier injection. Whether the former or the latter mechanism dominates, or whether there are some further extrinsic mechanisms at play (e.g. defects) that were not considered by Seidel et al. remains an open question that is left for future studies.

\section{Artificially layered ferroelectrics}
\label{s_artificiallayer}

An alternative approach to investigate and exploit the properties of 
ultrathin ferroelectrics has been to fabricate superlattices where 
ferroelectric and paraelectric layers as thin as a few unit cells 
alternate periodically to produce an artificially layered single 
crystal~\cite{Tabata-94}. 
These seemingly simple nominally two-component heterostructures have been 
found to display a rich spectrum of functionalities arising from the 
interplay between the effects of strain, electrostatic interactions 
between the ferroelectric layers, and coupling of different structural 
instabilities in the reduced symmetry environment of the interfaces.

\subsection{Electrostatic coupling}

When thin ferroelectric layers are separated by paraelectric slabs, 
any discontinuity in the polarization will give rise to strong 
electric fields that tend to both suppress the polarization in the 
ferroelectric layers and polarize the paraelectric component. 
The electrostatic energy cost associated with these fields is very large 
and thus the system will look for a more favorable ground state. 

\begin{figure}[!htb] 
   \begin{center}
      \includegraphics[width=10cm,keepaspectratio]{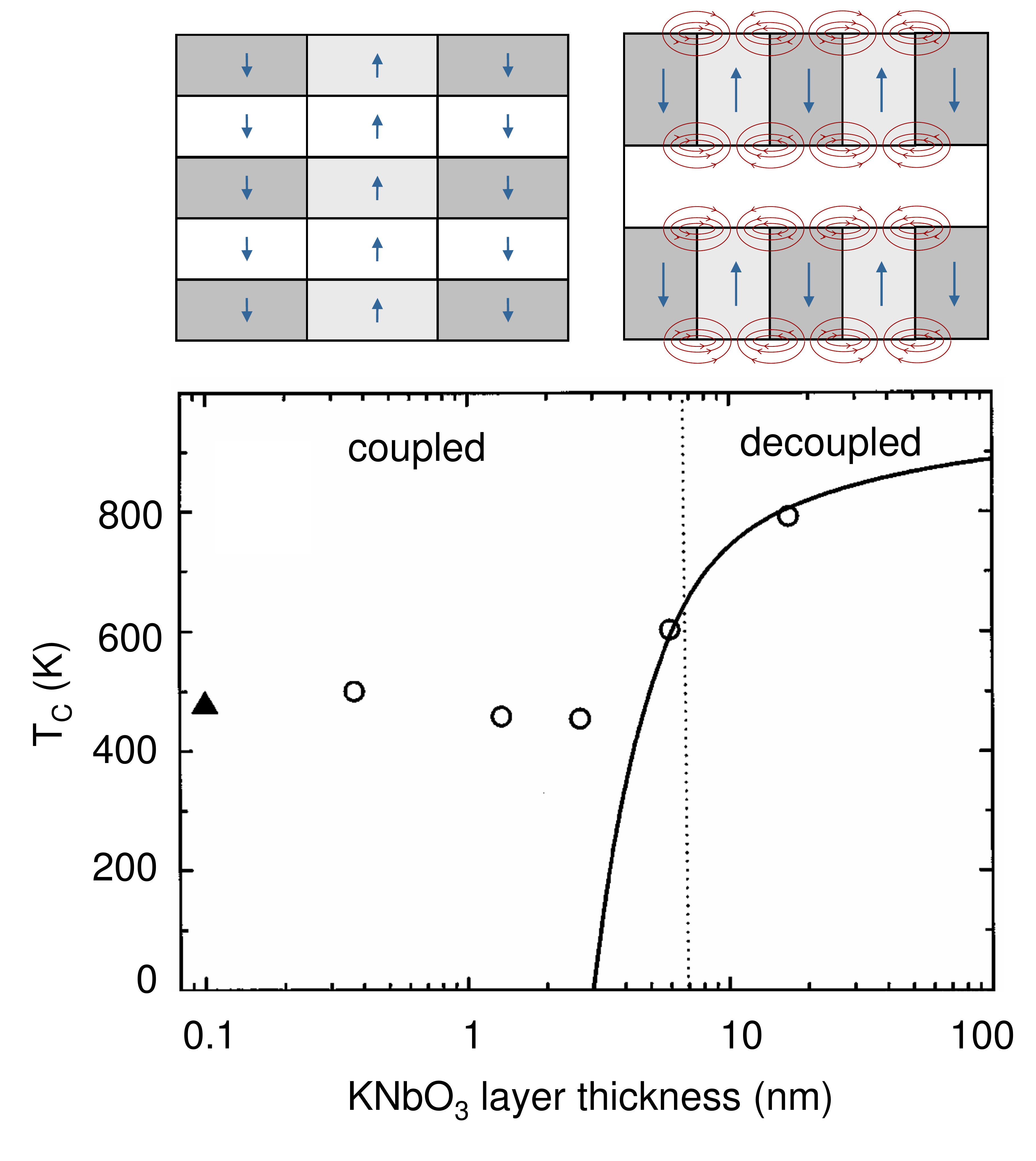}
      \caption{Evolution of the ferroelectric-to-paraelectric 
               transition temperatures with $n$ in 
               [KNbO$_3$]$_n$/ [KTaO$_3$]$_n$ superlattices, 
               illustrating a transition from strong interlayer
               coupling with continuous polarization thoughout the
               superlattice, to a decoupled state
               with polarization confined to ferroelectric
               layers~\cite{Specht-98,Stephanovich-05}.
               Top panels, after Fig. 1 of Ref.~\cite{Stephanovich-05}.
               Bottom panel adapted with permission 
               from Ref.~\cite{Specht-98}.}
      \label{f:coupling}
   \end{center}
\end{figure}

One possibility is to adopt a state of uniform out-of-plane 
polarization throughout 
the structure, as in Fig.~\ref{f:Screening}. In a first approximation, we can think in terms of a continuum model, based on the bulk electrical and electromechanical properties of the constituents. Within this assumption, the value of the polarization will only depend on the relative
fraction of the ferroelectric material and on the mechanical boundary
conditions~\cite{Neaton-03, Dawber-07}. 
For example, [BaTiO$_3$]$_1$/[SrTiO$_3$]$_4$ superlattices
with just a single unit cell of BaTiO$_3$ per period were predicted to be ferroelectric, 
whereas in [BaTiO$_3$]$_2$/[SrTiO$_3$]$_3$ superlattices with 2~u.c. 
of BaTiO$_3$ and 3~u.c of SrTiO$_3$ repeated periodically the polarization was even expected 
to exceed that of bulk BaTiO$_3$ due to the compressive strain imposed 
by the SrTiO$_3$ substrate~\cite{Neaton-03}. 
Experimentally, the enhancement of ferroelectricity in such fine-period 
two-component superlattices has been reported by a number of laboratories 
around the world, e.g. Refs~\cite{Shimuta-02, Tian-06, Tenne-06}. 
Similar behavior has also been observed in  three-component 
BaTiO$_3$/SrTiO$_3$/CaTiO$_3$ superlattices~\cite{Lee-05}. 
{\it Ab-initio} studies suggest that the intrinsic inversion 
symmetry breaking in such ``tricolor'' heterostructures should lead 
to a built-in bias and self poling~\cite{Sai-00}, 
with some experimental work supporting this
prediction~\cite{Warusawithana-03}.
As the individual layers get thicker, however, domain formation 
becomes a more effective mechanism for eliminating the depolarization fields 
and lowering the total energy of the superlattice~\cite{Lisenkov-07}. 
In [KNbO$_3$]$_n$/[KTaO$_3$]$_n$ superlattices, a crossover occurs 
when $n$ exceeds about 7~u.c. from a regime where the KNbO$_3$ layers  
are electrostatically coupled by the continuity of the 
displacement field, as described above, to a state where the 
polarization is confined mainly to the ferroelectric layers with domains 
of opposite polarization screening the depolarization 
field~\cite{Specht-98,Stephanovich-05,Sepliarsky-01}. 
Figure~\ref{f:coupling}(a) shows the ferroelectric-to-paraelectric 
transition temperatures as a function of $n$. The strong electrostatic 
interlayer coupling in short-period (small $n$) superlattices means they behave as 
a single ferroelectric material with $T_{\rm c}$ independent of $n$. 
In larger period superlattices the ferroelectric layers decouple 
and the transition temperature scales with the layer thickness as in 
single ferroelectric films. 
The densely packed nanodomains, characteristic of these ultrathin 
ferroelectric layers, give rise to a large enhancement in the 
effective dielectric response as even tiny domain wall displacements 
under applied field can lead to large changes in polarization~\cite{Zubko-10}. 

\begin{figure}[!htb] 
   \begin{center}
      \includegraphics[width=13cm,keepaspectratio]{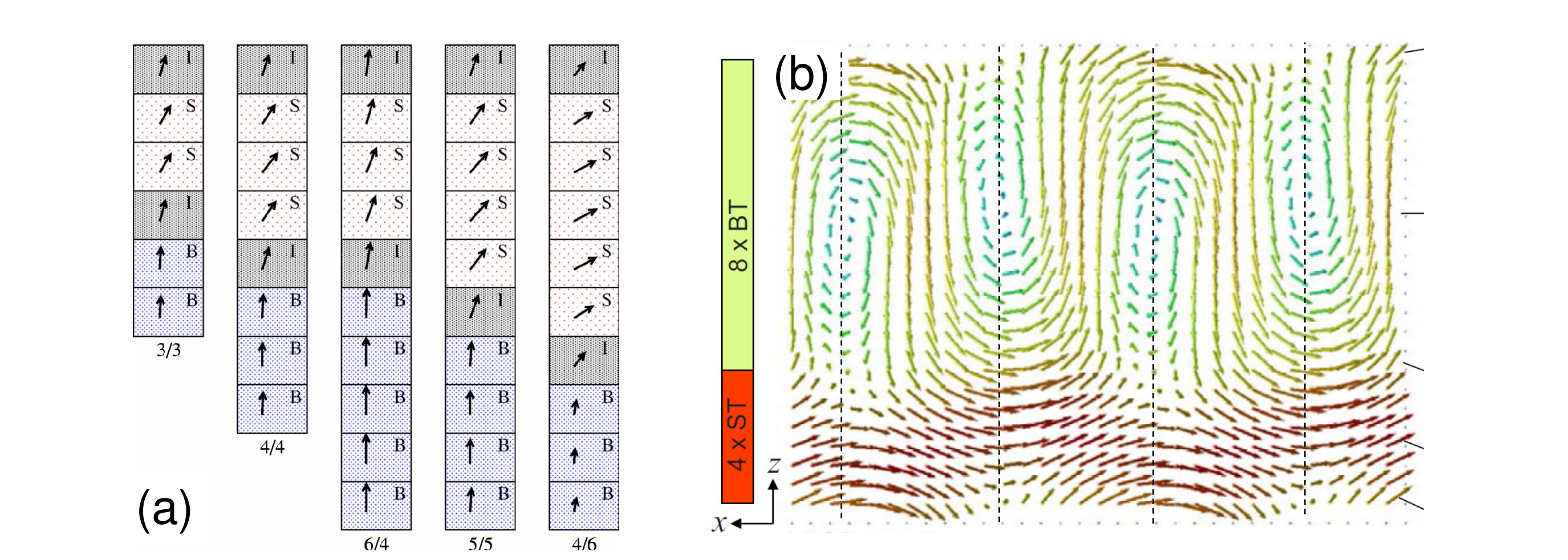}
      \caption {Theoretically calculated polarization distributions 
                in BaTiO$_3$/SrTiO$_3$ superlattices:
                (a) monodomain with different periodicities \cite{Johnston-05};
                (b) polydomain \cite{Lee-09}. 
                S, B and I in (a) denote SrTiO$_3$, BaTiO$_3$ and 
                interface layers respectively.  
                Reprinted with permission 
                from Ref.~\cite{Johnston-05} (panel a),
                and Ref.~\cite{Lee-09} (panel b).}
      \label{f:btosto}
   \end{center}
\end{figure}

Relaxing the substrate-imposed epitaxial strain leads 
to more complex behavior (Fig.~\ref{f:btosto}). In BaTiO$_3$/SrTiO$_3$ superlattices the 
BaTiO$_3$ layers will then impose a tensile strain on the SrTiO$_3$, 
forcing it into a strain-induced ferroelectric phase with 
in-plane polarization~\cite{Jiang-03,Johnston-05,Li-07}. 
The corresponding compressive strain experienced by the BaTiO$_3$ layers, 
on the other hand, reinforces the state with out-of-plane polarization. 
The interplay between mutual interlayer strains and the electrostatics 
demanding the continuity of the out-of-plane displacement leads to 
complex polarization rotation patterns and unusual domain structures, 
as shown in Fig.~\ref{f:btosto}~\cite{Johnston-05,Lee-09}. 
The morphologies and switching characteristics of domains in superlattices 
have already received  significant attention from the theoretical community, 
but remain challenging to study  experimentally. 

While simple arguments based on continuum electrostatics and elasticity theory are remarkably successful in describing many of the features of these nanoscale heterostructures, other properties require a more sophisticated theories. Recent studies highlight the importance of non-polar structural distortions, such as oxygen octaheral rotations as well as the properties of the interfaces between the individual layers \cite{Wu-08}. In short-period superlattices, genuine interfacial effects going beyond continuum electrostatic models become important. Quantifying these effects within first-principles theory became possible only recently, and the constrained D approach has been invaluable in decomposing the effects of electrostatics from those of short-range interactions~\cite{Stengel-09-np}. For example, in tricolor BaTiO$_3$/SrTiO$_3$/CaTiO$_3$ superlattices, Wu et al. \cite{Wu-08} found that interfaces tend to suppress ferroelectricity; that is, for a given relative fraction of the three constituents and a given value of the strain, P will tend to decrease for a higher density of interfaces. Interestingly, when octahedral rotations are allowed in the CaTiO$_3$ component of BaTiO$_3$/CaTiO$_3$ superlattices, interfaces appear to have a beneficial, rather than detrimental, effect~\cite{Wu-11}.

More generally, a major breakthrough in this field came from the realization that the broken-symmetry environment of the interface can have dramatic consequences on the electrical properties of the system, for example by ``activating'' lattice modes that are non-polar in the bulk phases of either constituent. We shall illustrate this important concept in the following section, by focusing on the case of oxygen octahedral rotations in PbTiO$_3$/SrTiO$_3$ superlattices.

\subsection{Engineering ferroelectricity at interfaces}

It has long been noted that the interface between two materials 
often has properties of its own, and in short-period superlattices, 
where the interface density becomes high enough, the interfaces can 
dominate the behavior of the material~\cite{Zubko-11}. 
Space charge accumulation and structural defects generally have 
detrimental effects~\cite{Oneill-00,Gregg-03}, and management of these 
extrinsic contributions is essential to uncover the more interesting 
intrinsic phenomena that arise at the interfaces. 

\begin{figure}[!htb] 
    \begin{center}
       \includegraphics[width=8cm,keepaspectratio]{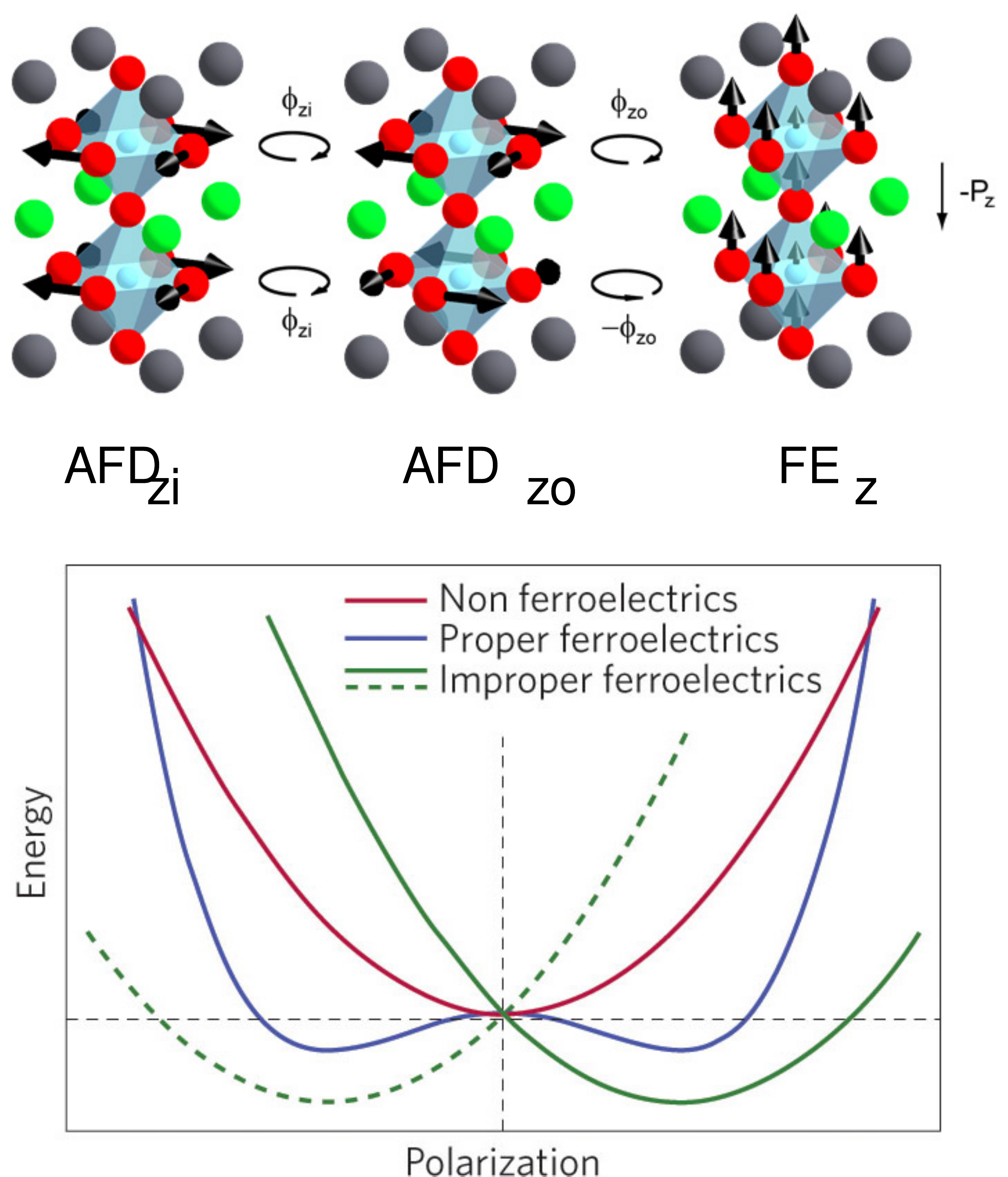}
       \caption {At the interface between PbTiO$_3$ and SrTiO$_3$,
                 the phase transition is driven by
                 the two antiferrodistortive modes that correspond to
                 in-phase (AFD$_{zi}$) and out-of-phase (AFD$_{zo}$)
                 rotations of the oxygen octahedra and induce the
                 polar distortion (P$_z$). 
                 The bottom panel illustrates, more generally,
                 how a ferroelectric state can be
                 induced in a non-ferroelectric material by either 
                 renormalizing
                 the curvature of the free energy to produce a double well
                 (e.g. by epitaxial strain) or by coupling to 
                 primary non-polar 
                 mode(s) inducing (hybrid) improper ferroelectricity
                 \cite{Ghosez-11}.
                 Reprinted with permission from Ref.~\cite{Ghosez-11}.} 
       \label{f:ptosto}
    \end{center}
 \end{figure}

The ideal cubic perovskite structure is stable only within a narrow 
range of ionic radii, and hence most ABO$_3$ perovskites are prone to 
a  number of symmetry lowering instabilities. These include polar 
distortions, oxygen octahedron rotations and Jahn-Teller distortions. 
In PbTiO$_3$, for instance, the polar and antiferrodistortive (AFD) 
instabilities compete and the condensation of the former suppresses 
the latter in bulk, whereas bulk SrTiO$_3$ is paraelectric but 
undergoes an AFD transition at around 105~K. While the different types of instabilities usually compete at the bulk level, ferroelectric (FE) and AFD modes can coexist at surfaces and interfaces, due to the asymmetric environment of the near surface atoms~\cite{Munkholm-02,Bickel-89}. In PbTiO$_3$/SrTiO$_3$ 1/1 superlattices epitaxially strained on SrTiO$_3$ substrates, it has been shown that the ground-state structure arises from the combined condensation of a FE distortion with amplitude $P_z$ and two distinct non-polar AFD motions involving rotations of oxygen octahedra either in-phase or out-of-phase by angles $\phi_{zi}$ and $\phi_{zo}$. These distortions are denoted respectively FE$_z$, AFD$_{zi}$ and AFD$_{zo}$ in Figure~\ref{f:ptosto}. Moreover, the symmetry relationship between these three distortions is such that it allows the appearance of an unusual trilinear coupling term, $−g\phi_{zi}\phi_{zo}P_z$, in the free energy expansion (Eq.\ref{eq:landau-expansion}), similar to what happens in improper ferroelectrics~\cite{Bousquet-08,Levanyuk-74}. In cases where the non-polar modes are primary order-parameters, the effect of this linear term on the free energy is to induce a spontaneous polarization by shifting the energy versus polarization well to lower energy as illustrated schematically in Figure~\ref{f:ptosto}. Improper ferroelectrics display a number of interesting and useful features such as temperature-independent dielectric permittivities~\cite{Levanyuk-74} and reduced sensitivity to depolarization-field-driven size effects~\cite{Sai-09}. 

Such trilinear coupling of structural instabilities at interfaces is not restricted to artificial PbTiO$_3$/SrTiO$_3$ superlattices and several recent theoretical studies have focused on the search for alternative materials presenting similar behavior. It has been demonstrated through group theoretical arguments, for example, that in bicolor ABO$_3$ superlattices a similar trilinear coupling between structural instabilities can only be achieved by modulating the A-site cation lattice (not the B-site). Moreover, such a coupling is not restricted to artificial superlattices but can also appear in naturally-occurring layered perovskites like Ruddlesden-Popper or Aurivillius compounds~\cite{Benedek-11}. The trilinear term produces an unusual coupling between FE and AFD modes in the sense that switching the polarization is mandatorily accompanied by the switching of one of the AFD order parameters. In magnetic materials, where magnetism arises from the superexchange interaction through the oxygens, such strong coupling between polarization and oxygen octahedron rotations is therefore a promising way of generating novel magnetoelectric couplings~\cite{Bousquet-08}. Ruddlesden–Popper compounds like Ca$_3$Mn$_2$O$_7$ have been recently proposed as possible candidates for realizing electric switching of the magnetization through so-called ``hybrid'' improper ferroelectricity~\cite{Benedek-11, Ghosez-11}, with the ferroelectric order parameter triggered by an effective field generated by the rotation of the oxygen octahedra.

\section{Conclusion and perspectives}
\label{s:conclusions}

Driven in part by the industrial needs for device miniaturization and in part by purely academic curiosity, the study of ferroelectrics is booming at the nanoscale. Materials processing as well as characterization techniques have finally reached the maturity  that allows us to experimentally probe fundamental sized effects rather than those imposed by processing limitations. Along with the experimental advances, one of the most significant milestones in the study of ferroelectricity has been the development of powerful first principles calculations. Not only have these shed enormous light on the fundamental origin of ferroelectricity, but they are also an ideal way for describing experimental findings and for understanding the nature of ferroelectric size effects.

In this chapter, we have summarized some of the recent developments in the theoretical and experimental understanding of ultrathin ferroelectrics. Many aspects of thin film ferroelectrics are dictated by the depolarization field and the way in which the material tries to minimize it. 
The fundamental discovery that, with proper management of the depolarization fields, there seems to be no intrinsic size limitations to ferroelectricity, has important implications for technological applications, promising further  downscaling of existing devices and opening doors to new devices based on ultrathin ferroelectrics. 
Studies of nanoscale domains and polarization vortices, as well as the properties of domain walls, have also been generating significant excitement, from both the theoretical and experimental communities, with many questions remaining unresolved and theoretical predictions to be confirmed.
An increasing amount of research is being focused on heterostructures involving ferroelectrics or materials that combine ferroelectricity with other interesting properties.  Today, there is tremendous excitement about the possibility of combining (and coupling) ferroelectricity and magnetism within a single multiferroic (magnetoelectric) compound or composite,  yielding truly multifunctional materials, novel device ideas and fascinating new physics. 

These recent experimental and theoretical developments also extend to the larger family of functional oxides, and fundamental research on this field, both on bulk phases and
on complex oxides heterostructures, is living a momentous stage. 
The driving force that fuels these efforts is based on the fact that,
within a relatively simple structure, these materials display a huge variety of 
properties that are susceptible to be used in electronic devices,
including high-T$_{\rm c}$ superconductivity
(with critical temperatures far higher than in standard superconductors),
colossal magnetoresistance (where the application of magnetic fields of a 
few teslas changes the electrical resistivity by orders of magnitude), 
ferroelectricity, magnetic ordering or even multiferroic behaviour.
These properties are the result of a subtle interplay between
different degrees of freedom, that include spin, charge, and orbital ordering.
Very interestingly, the energy scales of these interactions are of the
same order of magnitude, so a delicate balance between all of them
is always present. 
Small changes on the condition might incline the balance towards one of the
multiple local minima displayed in the energy landscape, yielding to 
very different ground states.

If already at the bulk level the physics of perovskite oxides is fascinating,
the combination of different materials to form superlattices and complex
oxide interfaces has opened new and unexpected research avenues.
The behaviour of these heterostructures is far from being a linear combination
of the properties of the two materials, as we've seen in the discussion of ferroelectric superlattices. Going beyond ferroelectric materials, the number of possible combinations is infinite, and completely exotic behaviours have already
been observed, including metallic states at the interface
between insulating materials or the appearance of improper ferroelectricity,
with the ferroelectric order parameter triggered by an effective field
generated by the rotation of the oxygen octahedra.
Besides, the superlattices offer a powerful method for 
engineering and tuning desired properties
by playing with the possible material combinations,
periodicity of the structure, the strain or the electrical 
boundary conditions.

With these new properties we can easily envisage some dream come true,
for instance, magnetically readable and electrically 
writable multiferroic devices, or systems with greatly enhanced thermoelectric
properties for energy harvesting. 

\subsection*{Acknowledgments}

This work was supported by the Swiss National Science Foundation through the National Center of Competence in Research, Materials with Novel Electronic Properties, 'MaNEP', division II, Oxides;  by the European Union FP7/2007-2013 project 'OxIDes' under Grant  n.228989; by the Interuniversity Attraction Poles Program of the Belgian State-Belgian Science Policy (Grant No. P6/42); by a Joint Research Action of the French Community of Belgium (ARC-TheMoTherm project); and by The Leverhulme Trust "International Network on Nanoscale Ferroelectrics".


\end{document}